\documentclass[conference]{IEEEtran}

\usepackage{cite}
\usepackage{amsmath}
\usepackage{hyperref}
\usepackage{hyperxmp}
\usepackage{graphicx}
\usepackage{textcomp}
\usepackage{xcolor}
\usepackage{multirow}
\usepackage{ltxtable}
\usepackage{soul}
\usepackage{tikz}
\usepackage{algorithm}
\usepackage{algpseudocode}
\usepackage{colortbl}
\usepackage{bm}
\usepackage{float}
\usepackage{longtable}
\usepackage{cellspace}
\usepackage{subfigure}
\usepackage{subcaption}
\usepackage{mdframed}
\usepackage{comment}
\usepackage{balance}
\usepackage{hyperref}
\newcommand{\cmmnt}[1]{}  

\def\BibTeX{{\rm B\kern-.05em{\sc i\kern-.025em b}\kern-.08em
    T\kern-.1667em\lower.7ex\hbox{E}\kern-.125emX}}

\begin{document}

\title{PIM-LLM: A High-Throughput Hybrid PIM Architecture for 1-bit LLMs}

\author{
    \IEEEauthorblockN{Jinendra Malekar, Peyton Chandarana, Md Hasibul Amin, Mohammed E. Elbtity, and Ramtin Zand}
    \IEEEauthorblockA{Computer Science and Engineering, University of South Carolina, Columbia, SC 29201\\
    Email: jmalekar@email.sc.edu, psc@email.sc.edu, ma77@email.sc.edu, elbtity@ieee.org, ramtin@cse.sc.edu}
}

\maketitle

\pagestyle{plain}

\begin{abstract}

In this paper, we propose PIM-LLM, a hybrid architecture developed to accelerate 1-bit large language models (LLMs). PIM-LLM leverages analog processing-in-memory (PIM) architectures and digital systolic arrays to accelerate low-precision matrix multiplication (MatMul) operations in projection layers and high-precision MatMul operations in attention heads of 1-bit LLMs, respectively. Our design achieves up to roughly 80$\times$ improvement in tokens per second and a 70\% increase in tokens per joule compared to conventional hardware accelerators. Additionally, PIM-LLM outperforms previous PIM-based LLM accelerators, setting a new benchmark with at least 2$\times$  and 5$\times$ improvement in GOPS and GOPS/W, respectively.



\end{abstract}


\section{Introduction}


Generative large language models (LLMs) using decoder-only transformers, such as GPT \cite{GPT}, OPT \cite{zhang2022opt}, and LLaMA \cite{touvron2023llama}, have garnered significant attention for their impressive performance across diverse tasks, from machine translation to code generation \cite{xu2024contrastive,ugare2024improving}. However, this remarkable performance has come with rising computational and energy costs \cite{stojkovic2024towards,reidy2023work}. As a result, recent research has focused on reducing the energy footprint of LLMs through model compression techniques like quantization \cite{xiao2023smoothquant,shao2023omniquant,malekar2024matmulmatmulera1bit,you2024shiftaddllm}. Recently, 1-bit LLMs have been introduced that use extreme quantization by employing binary/ternary weights \cite{wang2023bitnet, ma2024era}. 1-bit LLMs can achieve substantial performance improvements if specific hardware architectures and systems are designed to support binary and ternary matrix multiplication (MatMul) operations. 



\begin{figure}[]
\centering
\includegraphics[width=0.98\columnwidth]{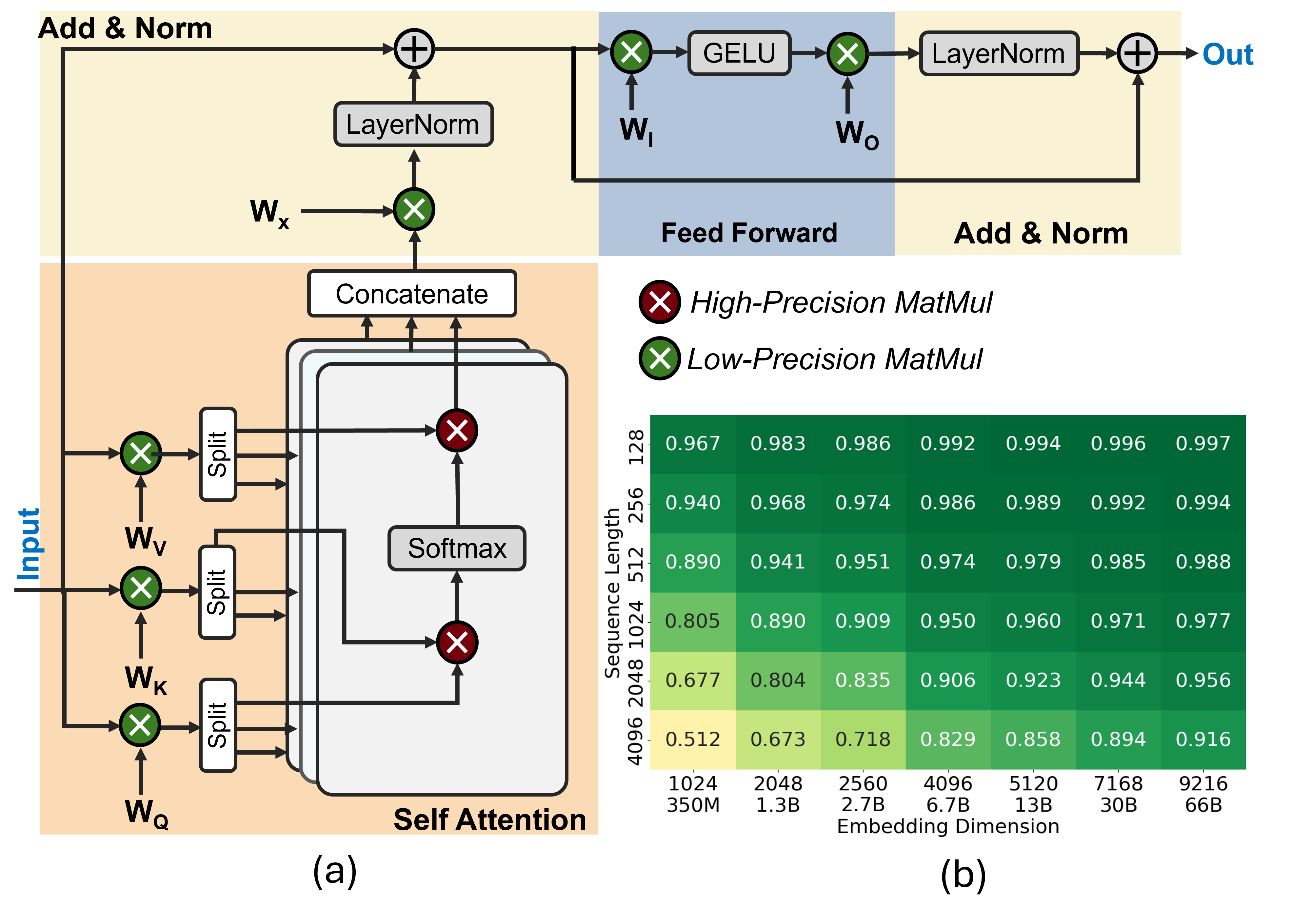}
\vspace{-3mm}
\caption{(a) The 1-bit LLMs divide the model into two portions: attention heads with high-precision MatMul operations (shown in red) and projection layers with low-precision MatMuls (shown in green). (b) The percentage of the low-precision MatMul operations in various OPT models.}
\label{fig:motivation}
\vspace{-6mm}
\end{figure}

It is essential to recognize that not all MatMul operations in 1-bit LLMs can undergo extreme quantization due to the resulting decrease in accuracy. Specifically, MatMul operations in the attention heads require higher precision, such as 8-bit integer. This divides the LLM models into two parts: one utilizing low-precision MatMul and the other relying on higher-precision MatMul, as illustrated in Fi. \ref{fig:motivation} (a). 
If the low-precision MatMul operations that are  in 1-bit LLMs account for most of the model's computational and memory demands, then focusing on custom hardware development to maximize the impact of extreme quantization would be worthwhile.

Figure \ref{fig:motivation} (b) shows the percentage of low-precision MatMul operations across various 1-bit OPT models \cite{zhang2022opt} with different sizes and context lengths. In all instances, the majority of computations are concentrated in the low-precision MatMul segment. The only case where the computation is more evenly distributed between low-precision and high-precision MatMuls occurs with the OPT 350M model at a 4096 context length. For larger models, the percentage of the low-precision MatMuls increases to more than 99\%, highlighting the potential performance and throughput benefits that can be attained by developing specialized hardware for executing binary and ternary MatMuls in 1-bit LLMs.

In this work, we introduce PIM-LLM, a hybrid architecture that integrates analog PIM with digital systolic arrays to efficiently execute low-precision and high-precision MatMuls in the projection layers and attention heads of 1-bit LLMs, respectively. Processing-in-memory (PIM) technology has been investigated as a beyond-von-Neumann architecture to accelerate machine learning (ML) workloads, particularly those with ternary and binary weights \cite{PIM-CNN, rakin2018pim}. PIM architectures commonly employ memristive crossbars to execute low-precision MatMul operations, leveraging massive parallelism and analog computation \cite{in-memory-dac}. Interest in using PIM technologies to accelerate language models has grown significantly in recent years \cite{PIM-GPT,TRANSPIM,iMCAT,RIME,ATT,ReBERT,ReTransformer,iMTransformer,X-Former,Hardsea,wolters2024memory}. However, many existing approaches primarily target encoder-only models or conventional encoder-decoder architectures, which differ in scale and computational demands from large generative language models. For example, designs like iMCAT \cite{iMCAT}, ATT \cite{ATT}, ReBERT \cite{ReBERT}, iMTransformer \cite{iMTransformer}, and X-Former \cite{X-Former} focus on smaller-scale, encoder-only models such as BERT variants. Meanwhile, RIME \cite{RIME} and ReTransformer \cite{ReTransformer} are PIM-based architectures designed to accelerate conventional encoder-decoder transformers \cite{vaswani2017attention}.

 This is the first study to specifically target the acceleration of modern decoder-only LLMs using analog PIM architectures, achieving state-of-the-art performance and throughput at a scale that has not been studied before.

\begin{figure}[t]
\centering
\includegraphics[width=\columnwidth]{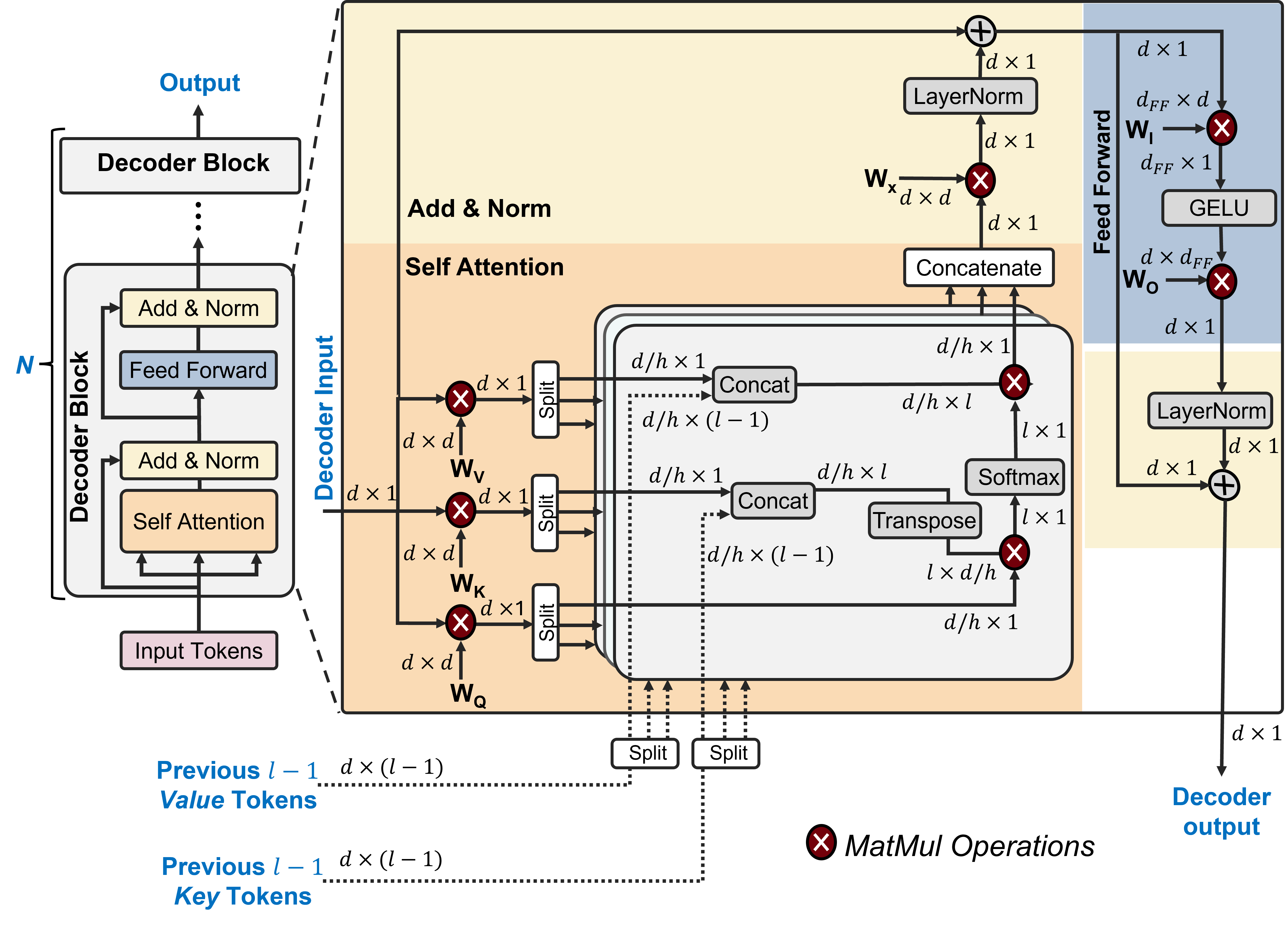}
\caption{The architecture of decoder-only LLMs. The tokenization and embedding layers are not shown in the figure.}
\label{fig:decoder}
\vspace{-3mm}
\end{figure}

\section{Underlying Operations of Decoder-Only LLMs}




Figure \ref{fig:decoder} shows the architecture of decoder-only LLMs, consisting of $N$ decoder blocks. Each block includes self-attention and feedforward layers followed by addition and normalization operations \cite{vaswani2017attention}. The self-attention mechanism starts with three linear projections of the token vector, $I$, into three distinct vectors: Key ($K=W_K.I$), Query ($Q=W_Q.I$), and Value ($V=W_V.I$), where $W_Q$, $W_K$, and $W_V$ are $d \times d$ weight matrices, in which $d$ is the embedding dimension.   




The $K$, $Q$, and $V$ vectors are then divided and distributed across $h$ attention heads. Next, each split value and key vector is concatenated with the previously cached $l-1$ value and key tokens from earlier token generation steps, creating a $d/h \times l$ matrix in each head, where $l$ represents the context length. The \textit{attention scores} are subsequently calculated using a scaled dot-product between the queries and keys, followed by multiplication with the generated value matrix. This process involves two MatMul operations denoted by ($.$) below:

\begin{equation}
    \text{Attention} (Q, K, V) = softmax (\frac{Q.K^T}{\sqrt{d}}) . V
\end{equation}

Next, the output of the attention heads are concatenated and transformed via another linear projection as follows:

\begin{equation}
    \text{MultiHead} (Q,K,V) = \text{Concat} (h_1, ..., h_h) . W_X
\end{equation}

\noindent where $h_i$ = Attention $(Q_i, K_i, V_i)$, and $W_X$ is a $d \times d$ weight matrix. Finally, the attention output passes through feed-forward (FF) layers, consisting of two MatMul operations and one nonlinear transformation.  As shown in Fig. \ref{fig:decoder}, Gaussian error linear unit (GELU) \cite{gelu} activation function is commonly used in LLMs to introduce nonlinearity. 








\begin{table}[]
\small
\centering
\caption{MatMul Operations in decoder-only LLMs.} 
\begin{tabular}{lcc}
\hline
Block                                                                           & Description     & Dimension                                                               \\ \hline
Att. Projections & $W_Q$, $W_K$, $W_V$, $W_X$         & $(d\times d). (d\times 1)$                                                                         \\ \hline
\multirow{2}{*}{\begin{tabular}[c]{@{}l@{}}Att. Head\end{tabular}}       & $Score=Q.K^T$             & $(l\times d/h) . (d/h\times 1)$            \\
                                                                                & $V.Score$         & $(d/h\times l) . (l\times 1)$               \\ \hline
\multirow{2}{*}{\begin{tabular}[c]{@{}c@{}}FF Projections\end{tabular}}                                                            & Intermediate FF & $(d_{FF} \times d) . (d\times 1)$            \\
                                                                                & Output FF       & $(d\times d_{FF}) . (d_{FF}\times 1)$         \\ \hline
\end{tabular}
\label{tab:dimension}
\end{table}

\begin{figure*}
    \centering
    \includegraphics[width=0.98\textwidth]{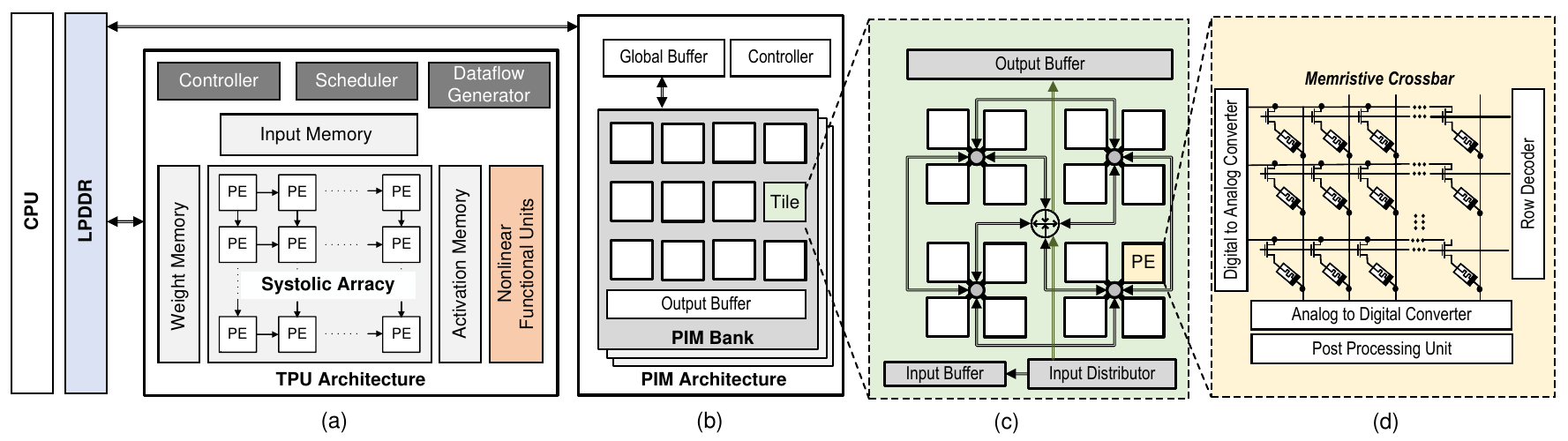}
    \caption{The proposed PIM-LLM architecture. (a) The LLM-specific TPU architecture, (b) The PIM architecture with multiple banks. (b) The PIM tile consists of a network of PEs. (c) The PEs include memristive crossbars to perform MVM operations.}
    \label{fig:arch}
    \vspace{-3mm}
\end{figure*}


Table \ref{tab:dimension} lists the dimensions of each MatMul operation in the decoder-only LLMs. In these models, MatMul operations effectively become matrix-vector multiplications (MVMs), as inference is performed iteratively, processing one input token per iteration while caching keys and values from previous iterations, as illustrated in Fig. \ref{fig:decoder}.  This can lead to substantial performance drops on traditional ML accelerators like TPUs, due to under-utilization of processing elements in their systolic arrays \cite{elbtity2023heterogeneous}. Similarly, PIM-based architectures, which are not tailored for decoder-only LLMs, face notable performance challenges when running these models \cite{Hardsea, TRANSPIM}.

Previous studies \cite{kim2023full} have demonstrated that by designing specialized hardware for nonlinear operations in LLMs (i.e., LayerNorm, Softmax, and GELU), the computational overhead from these operations can be made negligible compared to MatMuls. Therefore, focusing on optimizing MatMul operations is a more effective approach for achieving substantial improvements in LLM performance. The 1-bit LLMs \cite{wang2023bitnet,ma2024era,onebit} aim to accelerate the weight-to-activation MatMuls in the projection layers by extremely quantizing all weight matrices ($W_Q$, $W_K$, $W_V$, $W_X$, $W_I$, and $W_O$). However, activation-to-activation MatMuls within the attention heads still require high-precision MAC units. Fully unlocking the potential of such extreme quantization requires custom hardware. To this end, we propose a hybrid PIM-LLM architecture, designed to accelerate both binary/ternary MatMul operations in the projection layers and 8-bit MatMul operations in the attention heads, as described in the following section.

\section{Proposed PIM-LLM Architecture}

Here, we propose the hybrid PIM-LLM architecture consisting of components tailored to meet the implementation requirements of the projection layers and attention heads in 1-bit LLMs. These requirements include the need for a custom hardware to perform the low-precision W1A8 (1-bit weight, 8-bit activation) MatMuls in the projection layers. To accomplish this, we leverage analog PIM technology, employing memristive crossbars for 1-bit MatMul operations along with 8-bit analog-to-digital converters (ADCs) for generating the 8-bit activations. PIM architectures are typically used for weight-stationary dataflow, where weights are stored in RRAM crossbars, avoiding frequent crossbar updates. The activation-to-activation MatMuls, however, necessitate memory writes for each inference, resulting in substantial write energy overheads and  potential device failures due to the endurance limitations of memristive devices \cite{Endurance1}. Due to reliability and accuracy concerns, we do not use PIM technology for implementing the activation-to-activation MatMul operations in the attention heads. Instead, we use a digital systolic array architecture with 8-bit multiply-and-accumulate (MAC) units that can handle the W8A8 MatMuls in the attention heads.





\subsection{Design of LLM-Specific Tensor Processing Unit (TPU)}


Figure \ref{fig:arch} (a) illustrates the custom TPU architecture developed in this work to perform MatMul operations within the attention heads. This TPU design features dedicated memory units for weights, inputs, and outputs, along with a $32 \times 32$ systolic array. Additionally, it includes a specialized \textit{Nonlinear Functional Unit}, equipped with custom hardware, such as ConSmax \cite{liu2024consmax}, to execute the \textit{Softmax} operations in the attention heads of LLMs. The systolic array includes a two-dimensional grid of processing elements (PEs), where each PE executes a MAC operation. Specifically, it multiplies 8-bit weights and activations via an 8-bit multiplier circuit and adds the result to accumulated partial sums through an accumulator circuit.

\begin{figure}[t]
\centering
\includegraphics[width=0.95\columnwidth]{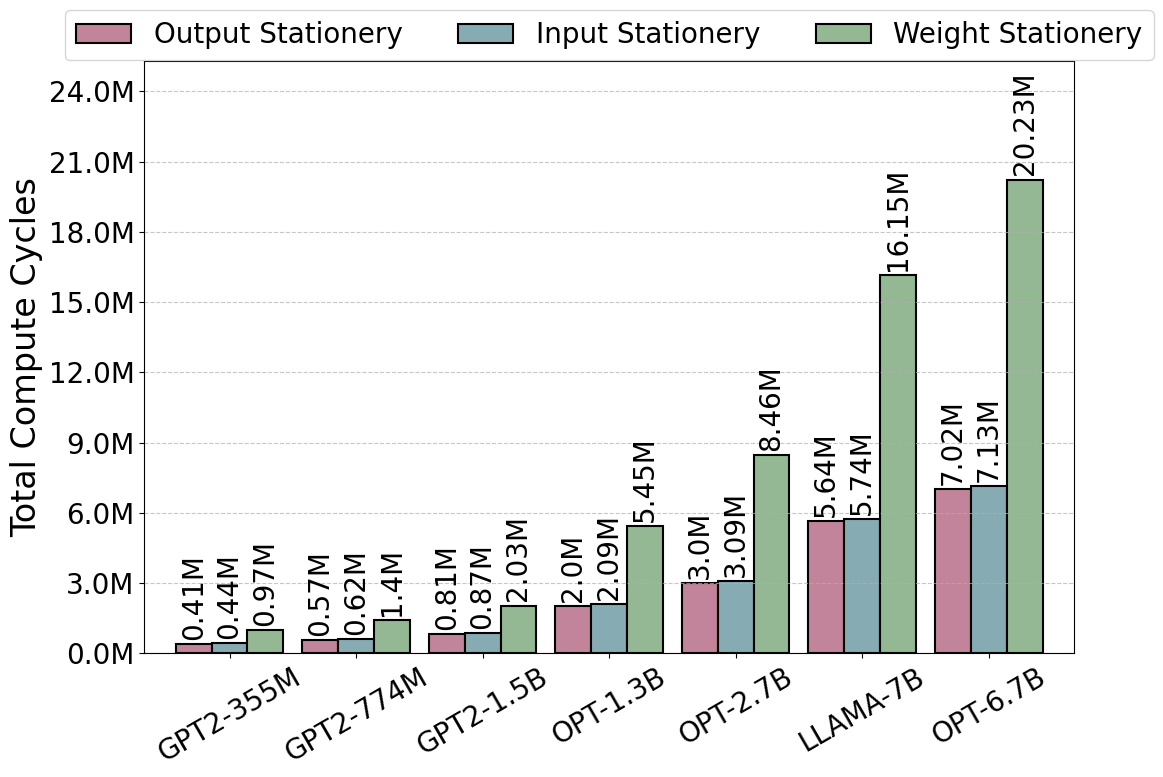}
\caption{Total cycles required for executing various LLMs using $32\times32$ systolic arrays with different dataflow architectures.}
\label{fig:dataflow}
\vspace{-3mm}
\end{figure} 

For activation-to-activation MatMuls in the attention head, the concatenated \textit{Value} and \textit{Key} matrices are stored in the weights memory, while the \textit{Query} and \textit{attention score} vectors reside in the input memory. For the dataflow architecture, we selected an output-stationary (OS) approach, which, based on our cycle-accurate analysis using the SCALE-Sim framework \cite{samajdar2018scale}, demonstrated better performance over weight-stationary (WS) and input-stationary (IS) dataflows (see Fig. \ref{fig:dataflow}). In the OS dataflow, inputs and weights are fetched from memory, multiplied, and the resulting products are accumulated with partial sums that remain stationary within the PEs \cite{APTPU}.


In the TPU architecture, the \textit{scheduler} orchestrates the execution of each layer in the workload, while the \textit{dataflow generator} and \textit{Main Controller} oversee the overall flow of the LLM workload. Depending on the workload requirements, the \textit{scheduler} may execute multiple layers sequentially. In the proposed PIM-LLM architecture, we use low-power double data rate (LPDDR) memory, which is well-suited for edge devices due to its low operating voltage and power-saving features. The data is preloaded into LPDDR, and as the TPU begins execution, the dataflow generator produces read address traces to retrieve inputs and weights from LPDDR, routing them to the input and weight SRAMs based on the OS dataflow algorithm. The main controller coordinates this data transfer according to the scheduler’s instructions.

\subsection{Design of LLM-specific PIM Architecture}


\begin{figure*}[t]
\centering
\subfigure[$l=128$]{\includegraphics[width=0.31\textwidth]{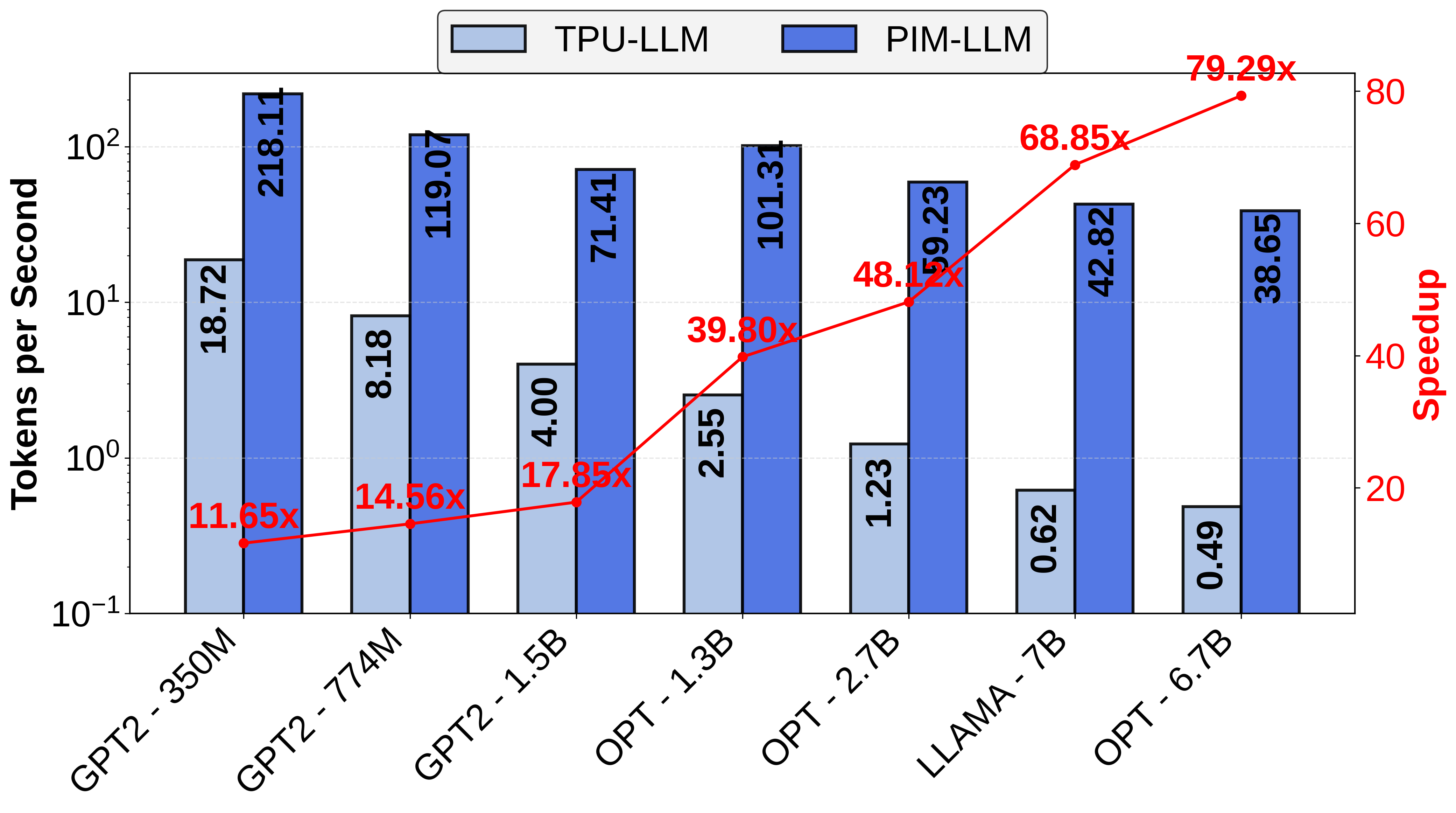}}
\hfill
\subfigure[$l=256$]{\includegraphics[width=0.31\textwidth]{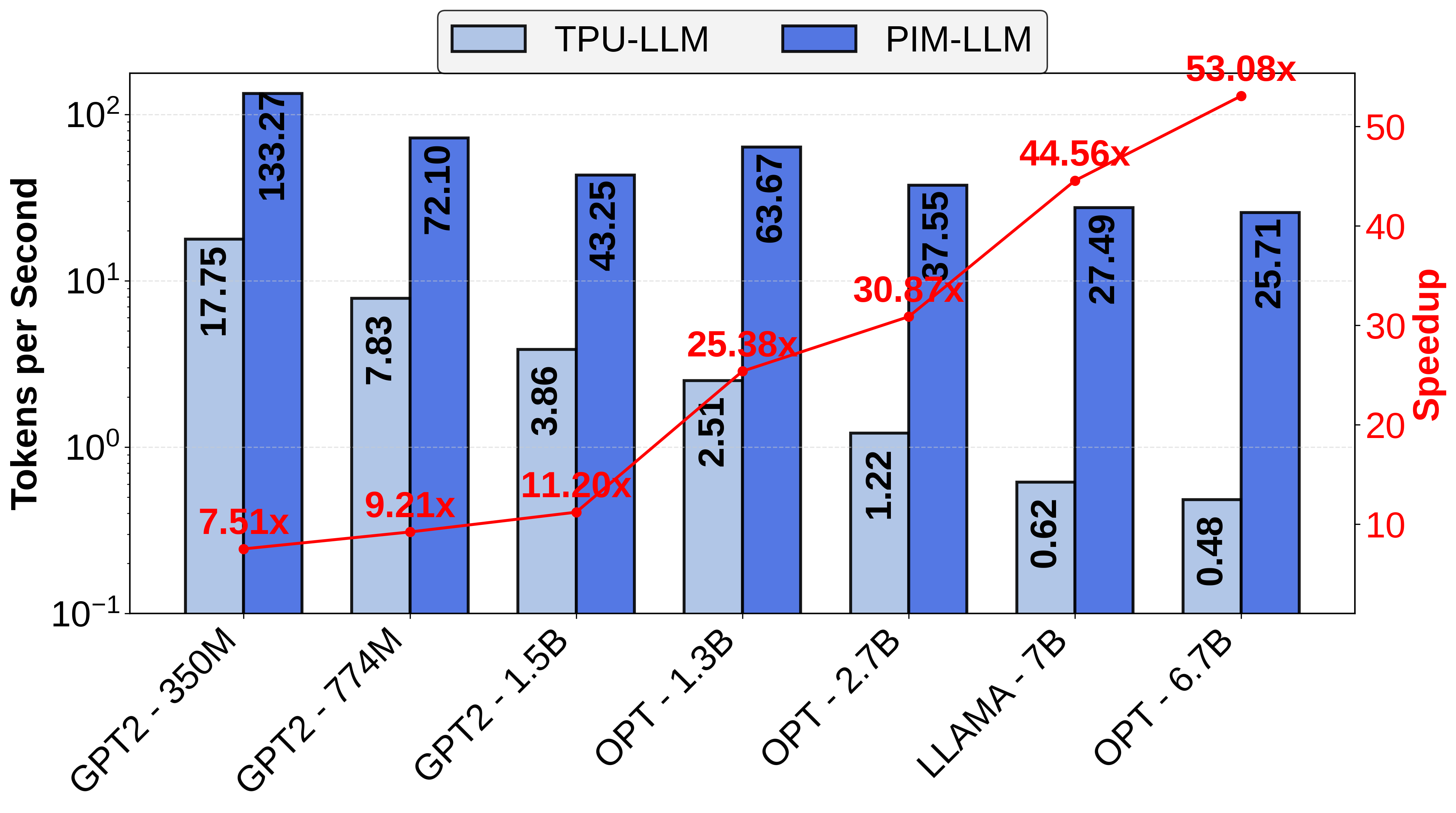}}
\hfill
\subfigure[$l=512$]{\includegraphics[width=0.31\textwidth]{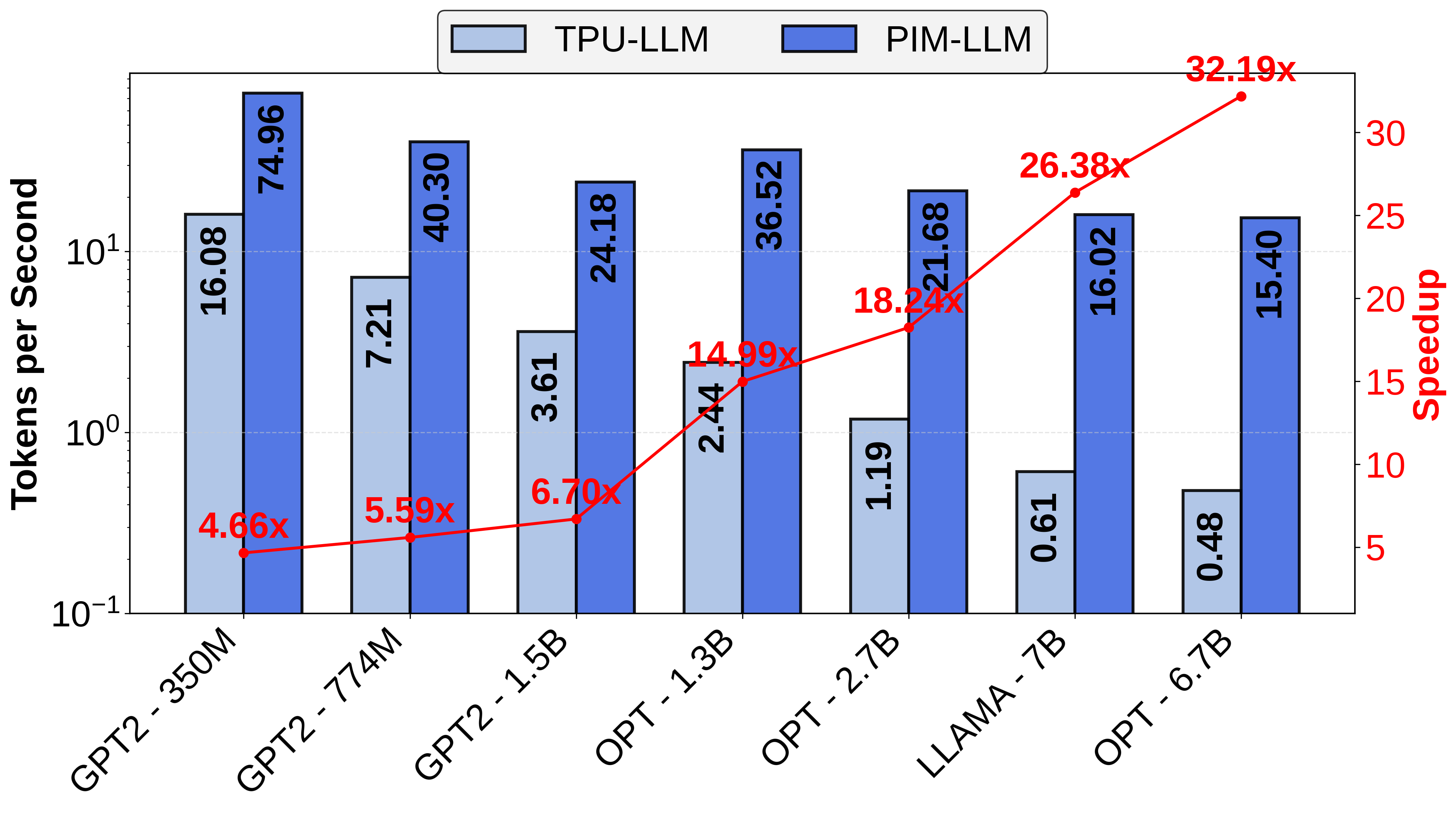}}
\hfill
\subfigure[$l=1024$]{\includegraphics[width=0.31\textwidth]{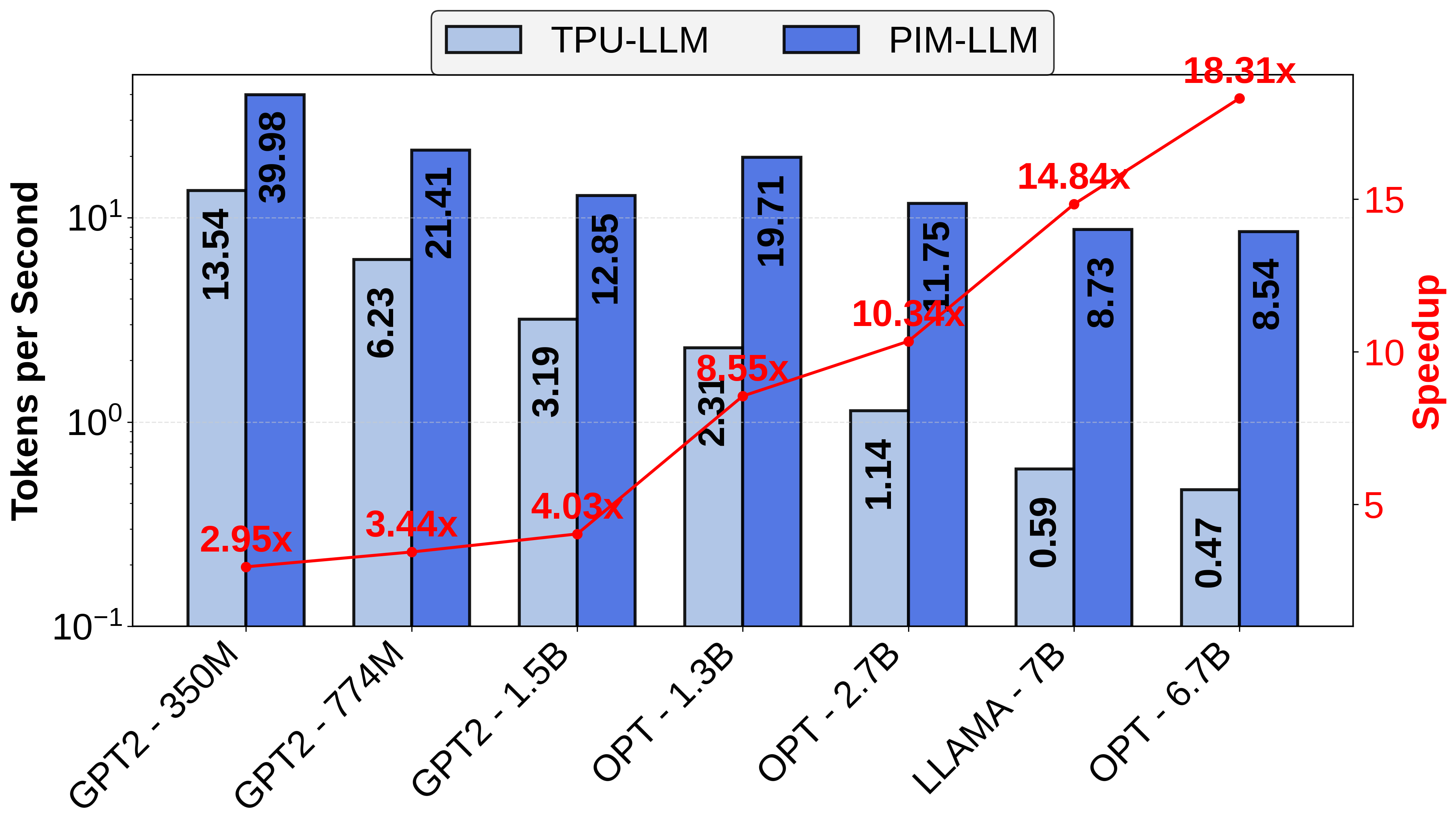}}
\hfill
\subfigure[$l=2048$]{\includegraphics[width=0.31\textwidth]{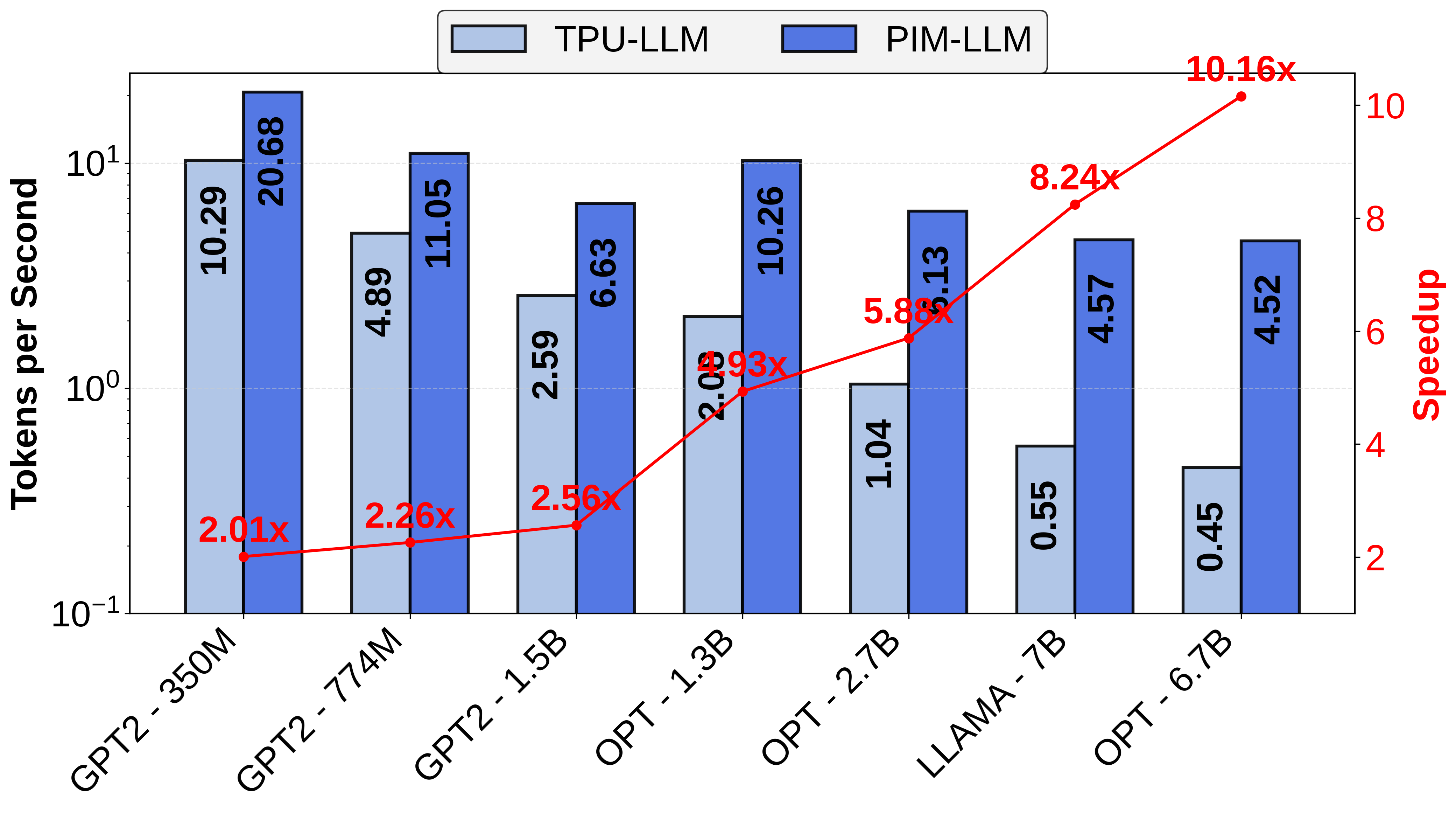}}
\hfill
\subfigure[$l=4096$]{\includegraphics[width=0.31\textwidth]{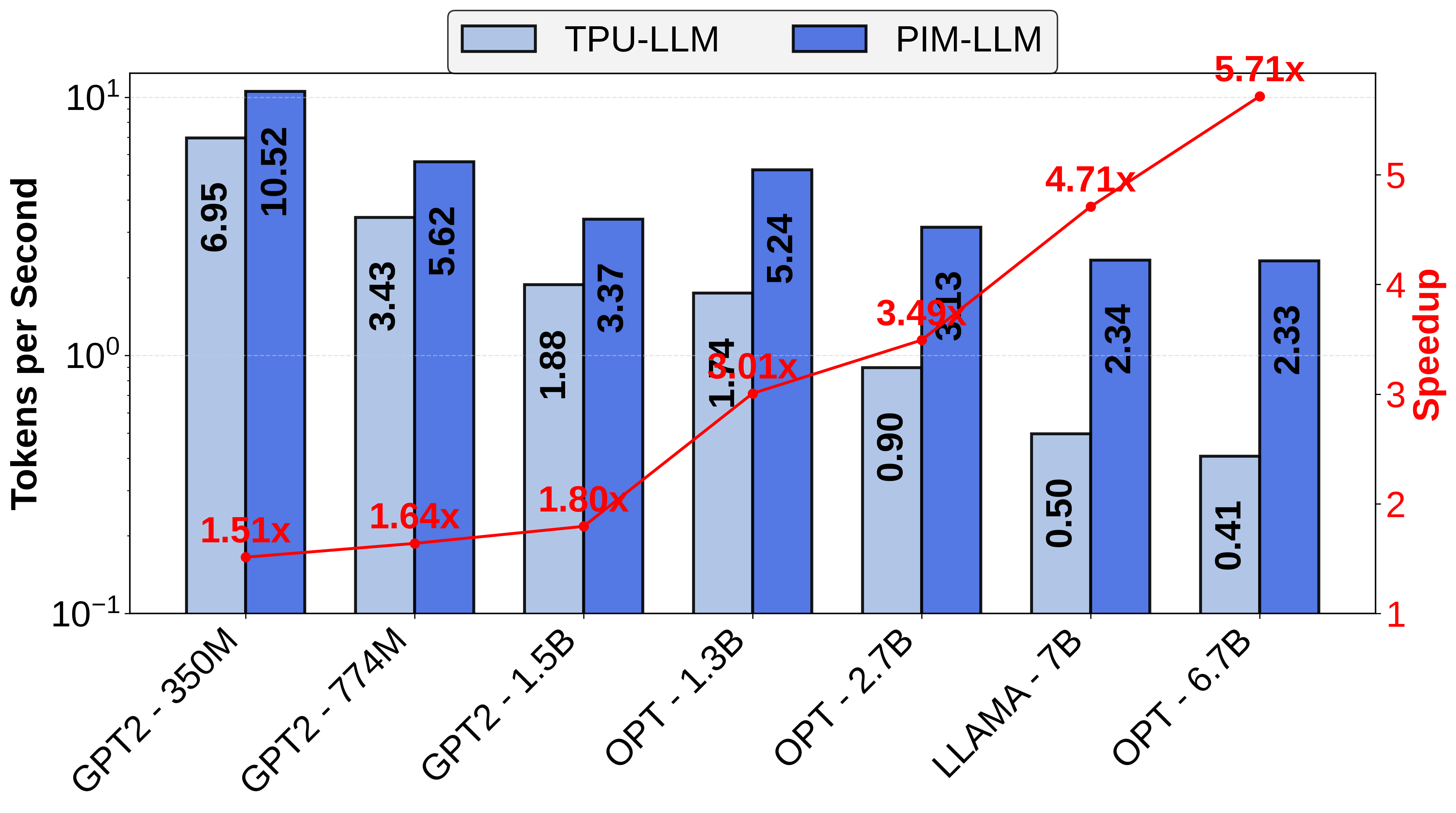}}
\hfill
\vspace{-3mm}
\caption{Tokens per second result for various LLMs with different context lengths ($l$).}
\label{fig:tokens_sec_speedup}
\vspace{-3mm}
\end{figure*}

Figure \ref{fig:arch} (b) show the overall PIM architecture, which includes multiple banks, a global buffer, and a controller. The PIM controller manages data movement between LPDDR and the PIM banks based on status updates from the CPU. Each PIM bank contains an array of tiles interconnected through a network-on-chip. Within each tile, multiple PEs, input buffers, and output buffers are arranged, as depicted in Fig. \ref{fig:arch} (c). Each PE block consists of memristive crossbars, digital-to-analog converters (DACs), analog-to-digital converters (ADCs), and postprocessing units that performs LayerNorm and GELU operations in the projection layers of LLMs.


Memristive crossbars, shown in Fig. \ref{fig:arch} (d), perform binary/ternary MVM operations in the analog domain using pairs of memristive devices and differential amplifiers \cite{elbtity2023heterogeneous,PIM-CNN,amin2022mram}. In the PIM architecture, projection layer weights are pre-loaded onto the memristive devices in the PIM banks during configuration. The weight kernels are expanded into vectors and loaded onto the crossbar columns, with input/activation vectors provided as inputs. The crossbars carry out MVM operations in parallel, applying Kirchhoff's and Ohm's Laws for analog computation \cite{in-memory-dac,PIM-CNN}. Once the projection layer computations are complete, results are digitized using ADCs and stored back in LPDDR for user access.

\section{Experiments and Discussions}

In this work, we designed the digital TPU architecture using Verilog and synthesized it with Synopsys Design Compiler at 45nm process technology to evaluate power and timing. For a $32 \times 32$ systolic array with 8-bit MAC units within its PEs, an operating frequency of 100 MHz was achieved. The TPU architecture includes 8 MB of SRAM, typical for edge TPU accelerators \cite{reidy2023efficient}. To simulate the PIM component and assess its latency and energy consumption, we used MNSIM 2.0 \cite{MNSIM2} with 256 $\times$ 256 RRAM crossbars and 45nm 8-bit ADCs \cite{choi2015ADC}. We start our experiments by benchmarking the throughput of our hybrid PIM-LLM architecture against the baseline LLM-specific TPU (TPU-LLM) design using various LLMs listed in Table \ref{tab:modelhyperparameters}. Finally, we compare our architecture with previous PIM-based designs developed to accelerate language models.



\begin{table}[]
\centering
\caption{The hyper-parameters of LLMs.}
\vspace{-2mm}
\begin{tabular}{c|c|ccccc}
\hline
\multirow{2}{*}{Models} & \multirow{2}{*}{Param} & \multicolumn{5}{c}{Hyperparameters}                                                              \\ \cline{3-7} 
                        &                        & \multicolumn{1}{c}{$d$}    & \multicolumn{1}{c}{$h$}        & \multicolumn{1}{c}{$d_{FF}$}   & $l$   & $N$  \\ \hline
\multirow{3}{*}{GPT}    
                        & 355M                   & \multicolumn{1}{c}{1024} & \multicolumn{1}{c}{16}       & \multicolumn{1}{c}{1024}  & 128-4096 & 24 \\ 
                        & 774M                   & \multicolumn{1}{c}{1280} & \multicolumn{1}{c}{20}       & \multicolumn{1}{c}{1280}  & 128-4096 & 36 \\ 
                        & 1.5B                   & \multicolumn{1}{c}{1600} & \multicolumn{1}{c}{25}       & \multicolumn{1}{c}{1600}  & 128-4096 & 48 \\ \hline 
\multirow{3}{*}{OPT}    
& 1.3B                   & \multicolumn{1}{c}{2048} & \multicolumn{1}{c}{32}       & \multicolumn{1}{c}{8192}  & 128-4096 & 24 \\ 
                        & 2.7B                   & \multicolumn{1}{c}{2560} & \multicolumn{1}{c}{32}       & \multicolumn{1}{c}{10240} & 128-4096 & 32 \\ 
                        & 6.7B                   & \multicolumn{1}{c}{4096} & \multicolumn{1}{c}{32}       & \multicolumn{1}{c}{16384} & 128-4096 & 32 \\ 
                         \hline
\multirow{1}{*}{LLaMA}  & 7B                     & \multicolumn{1}{c}{4096} & \multicolumn{1}{c}{32}       & \multicolumn{1}{c}{11008} & 128-4096 & 32 \\ 
                         \hline
\end{tabular}
\label{tab:modelhyperparameters}
\end{table}

\begin{figure}[t]
\centering
\subfigure[$l=128$]{\includegraphics[width=0.48\textwidth]{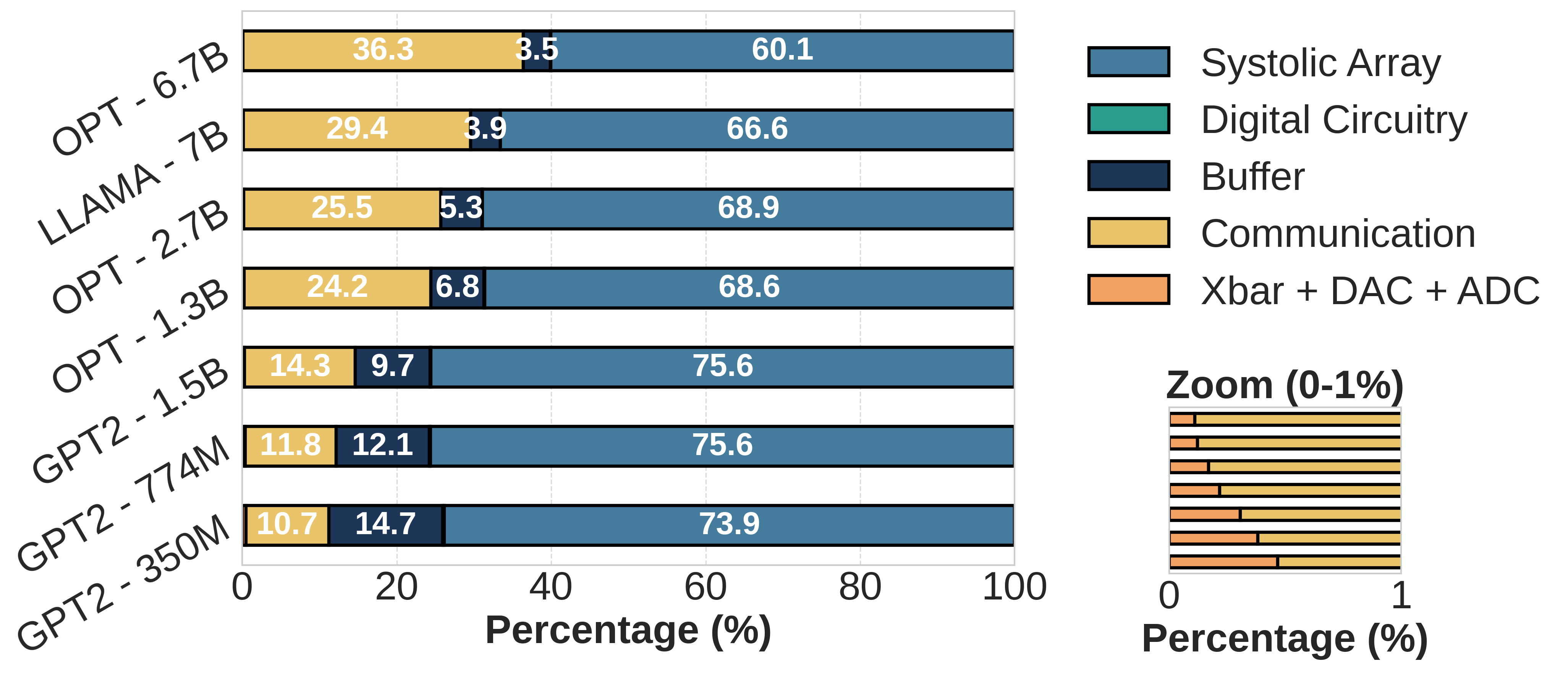}} \vspace{-2mm}
\subfigure[$l=4096$]{\includegraphics[width=0.48\textwidth]{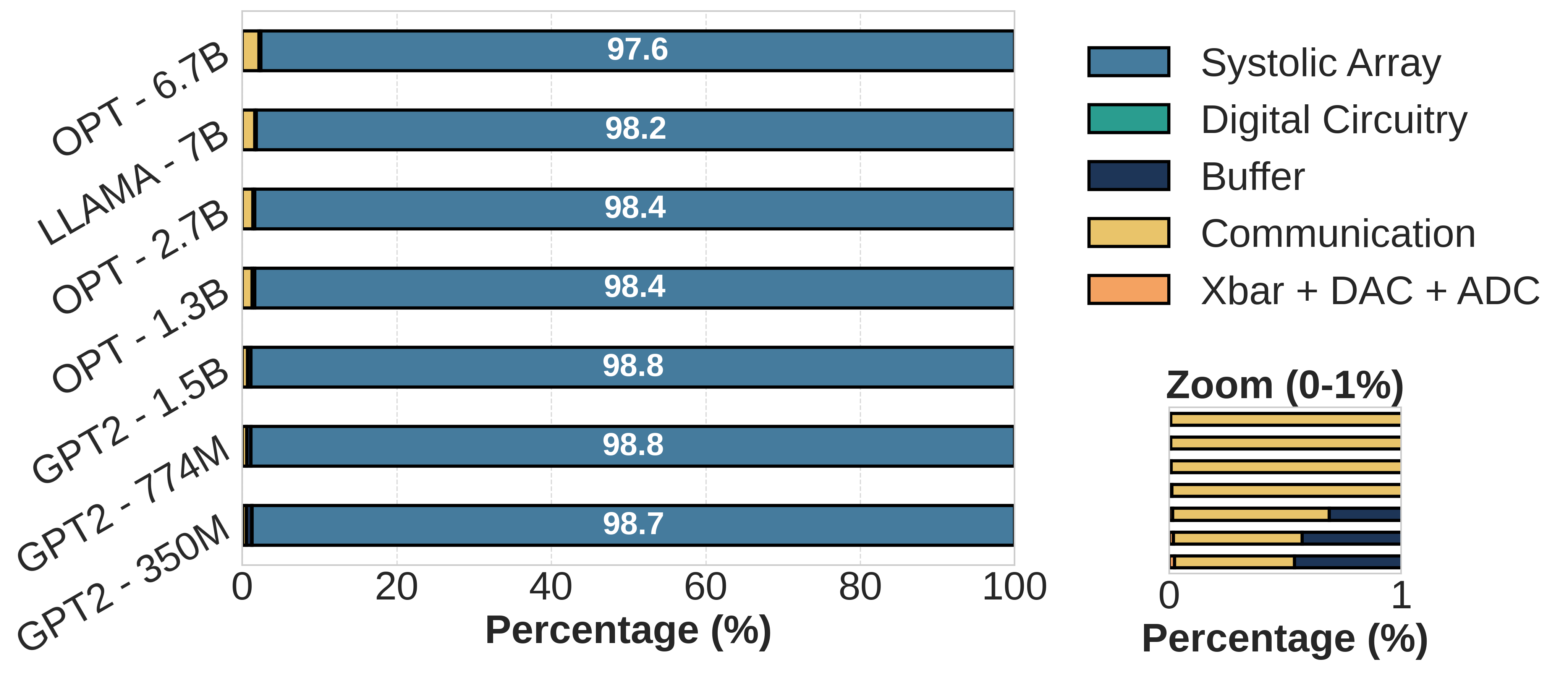}}
\hfill
\caption{Contribution of various components to the overall latency in the PIM-LLM architecture (percentage breakdown).}
\label{fig:latency}
\end{figure}

\begin{figure*}[t]
\centering
\subfigure[$l=128$]{\includegraphics[width=0.31\textwidth]{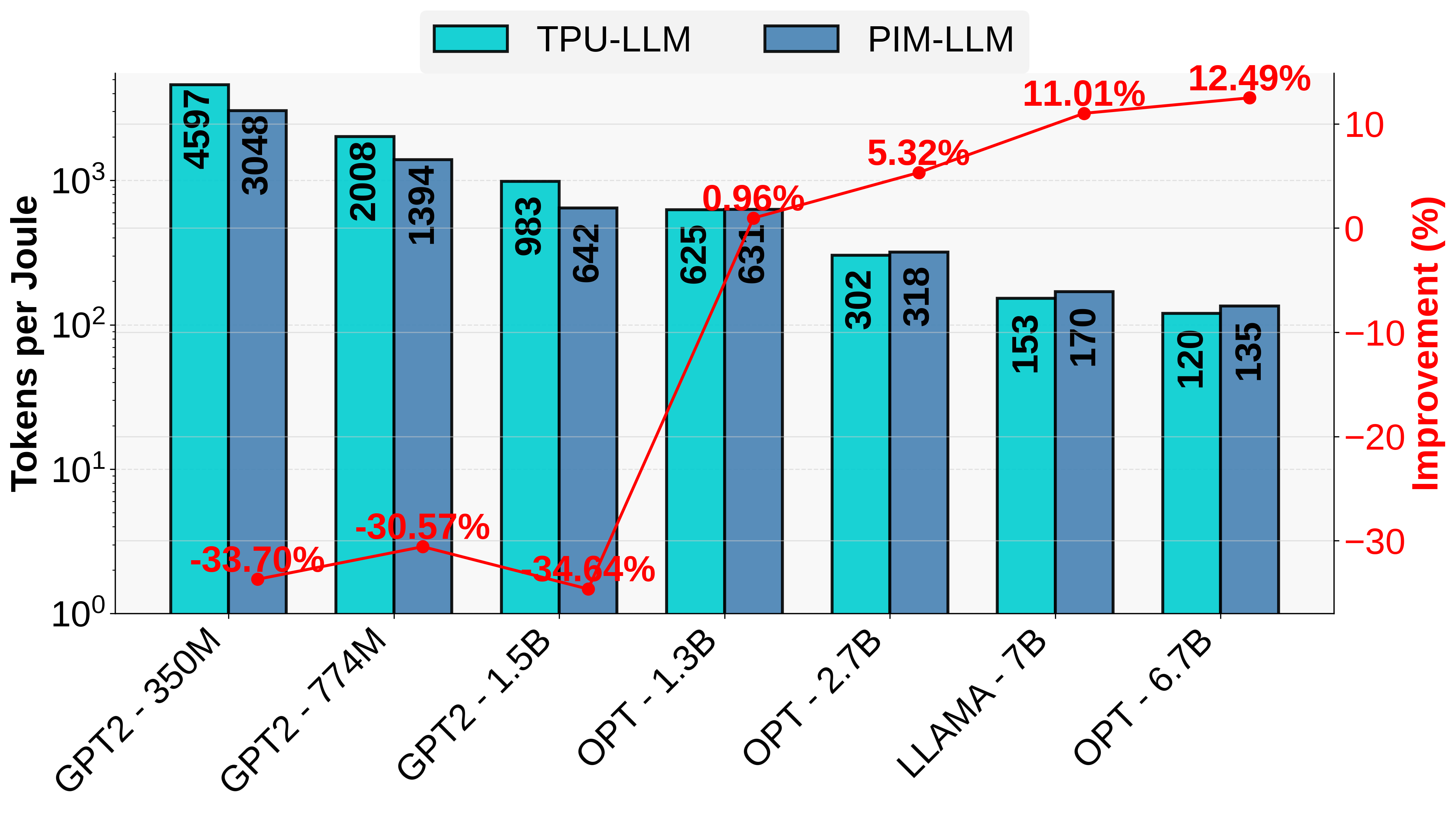}}
\hfill
\subfigure[$l=256$]{\includegraphics[width=0.31\textwidth]{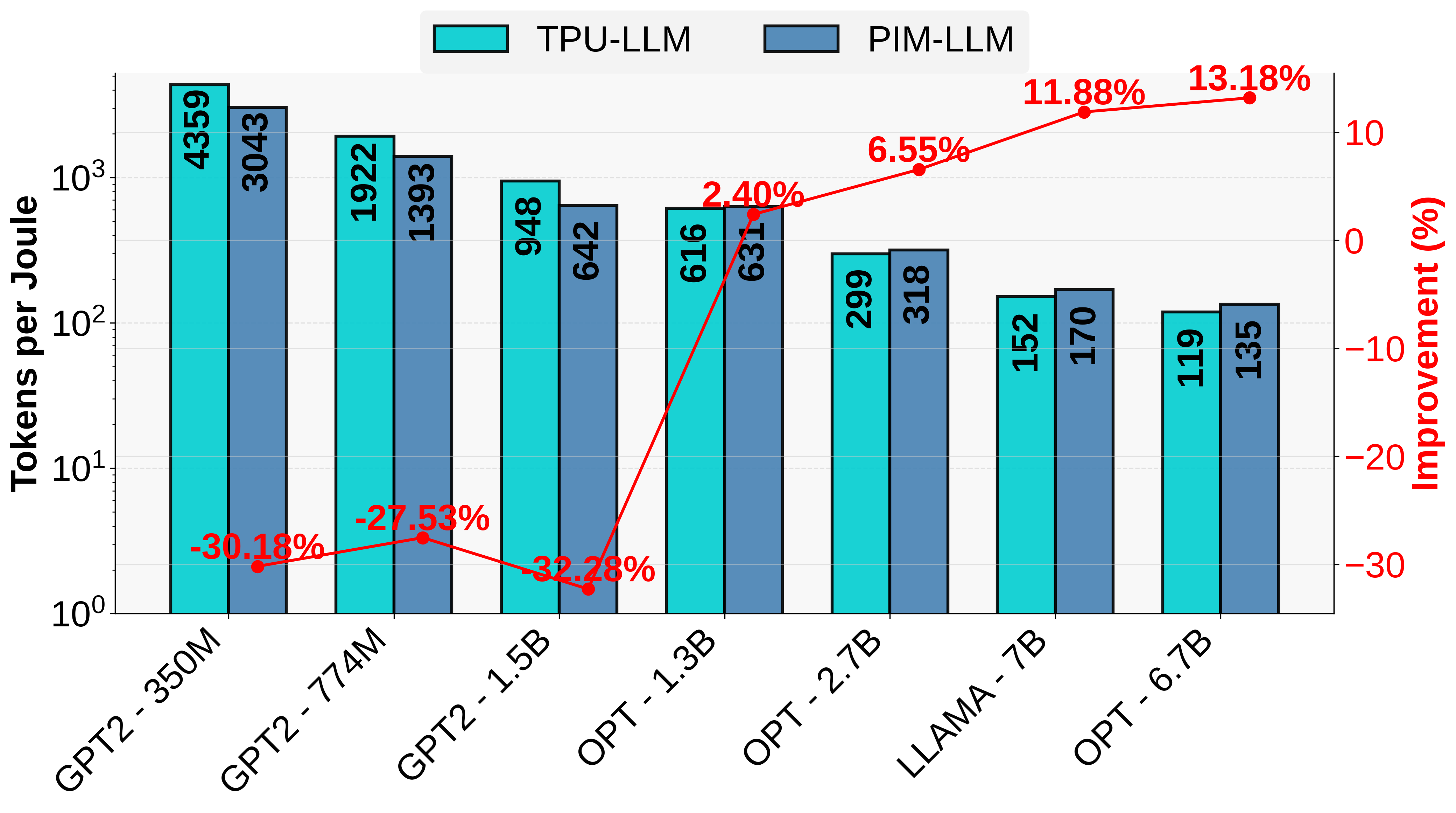}}
\hfill
\subfigure[$l=512$]{\includegraphics[width=0.31\textwidth]{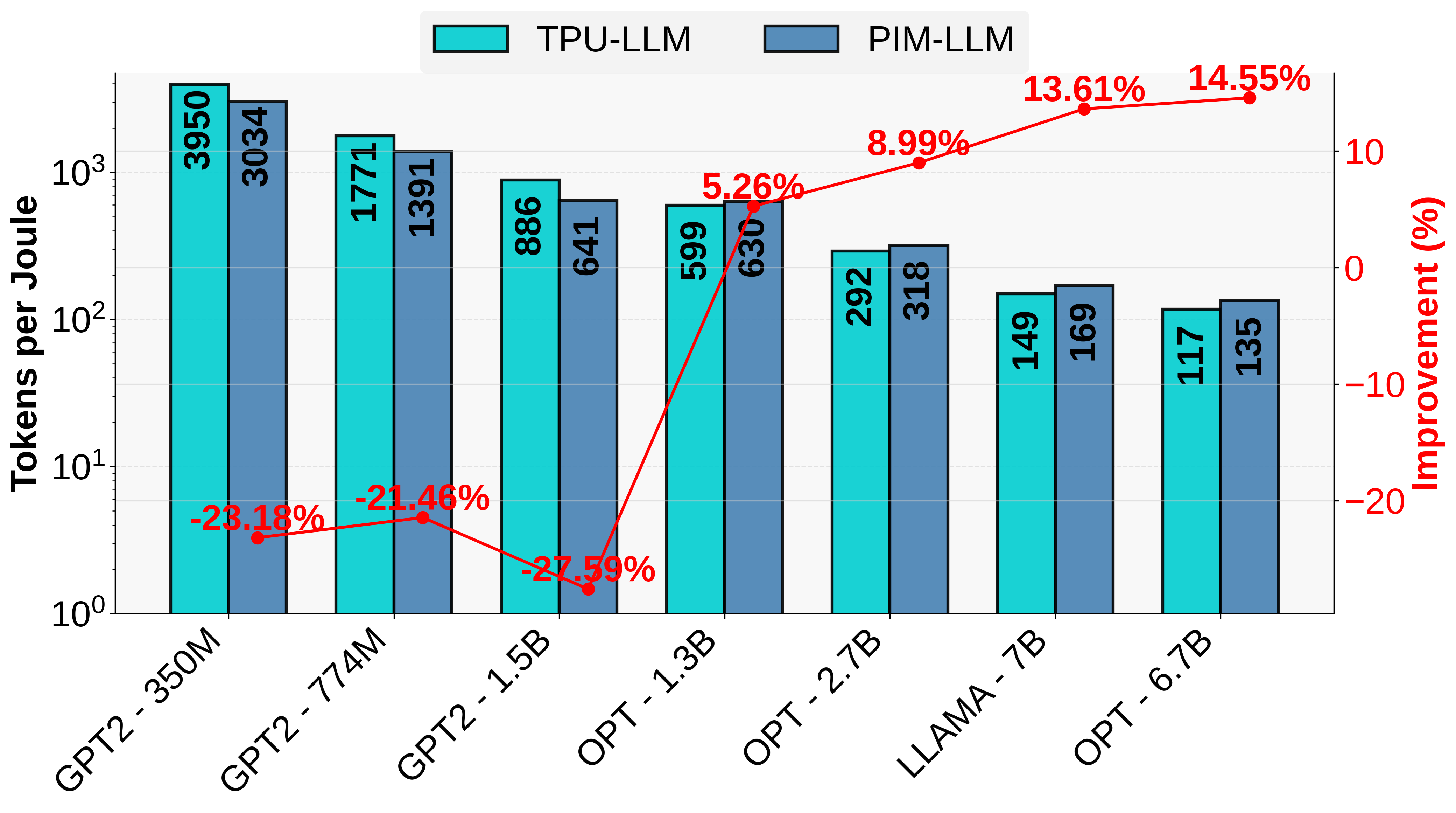}}
\hfill
\subfigure[$l=1024$]{\includegraphics[width=0.31\textwidth]{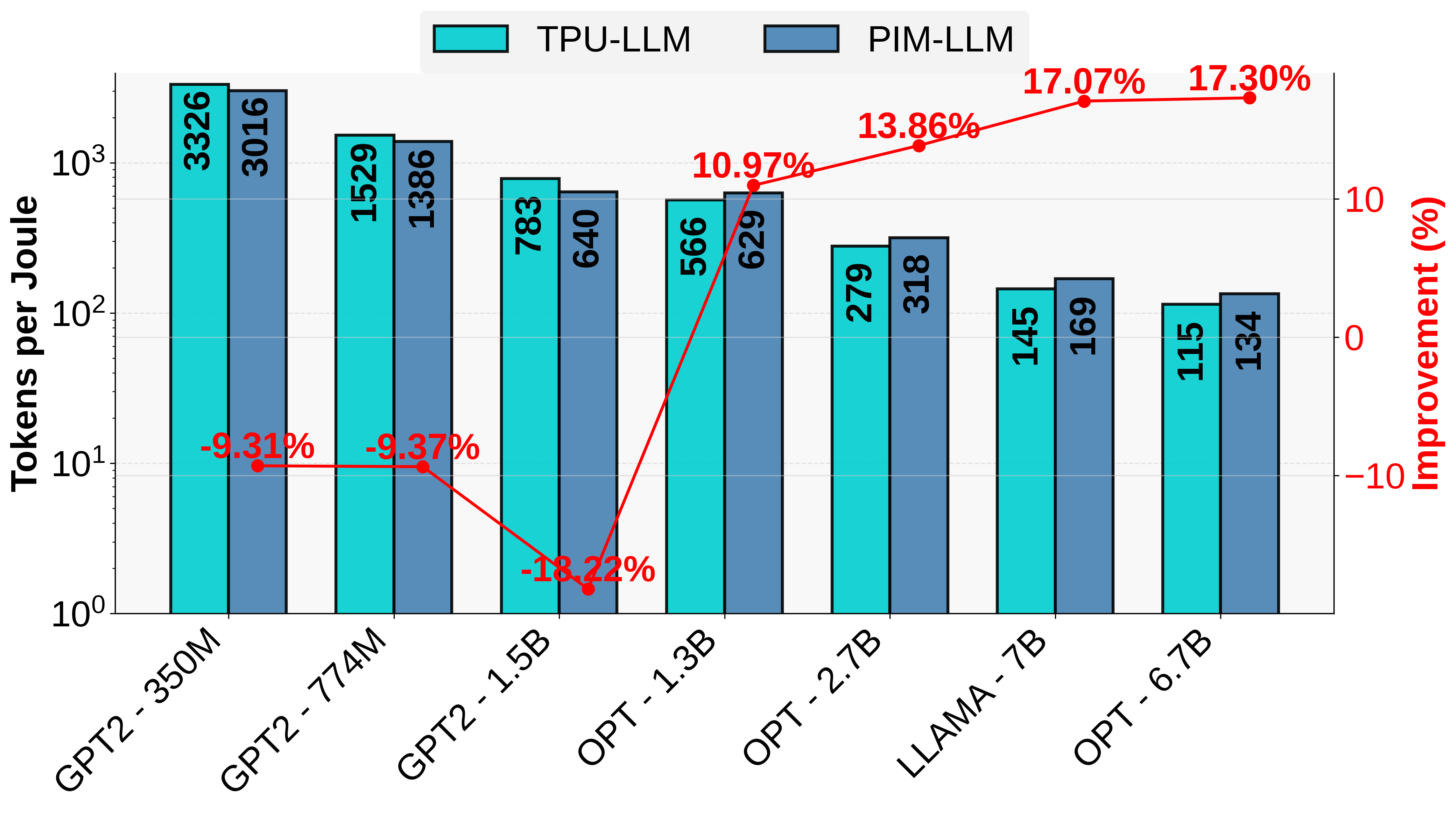}}
\hfill
\subfigure[$l=2048$]{\includegraphics[width=0.31\textwidth]{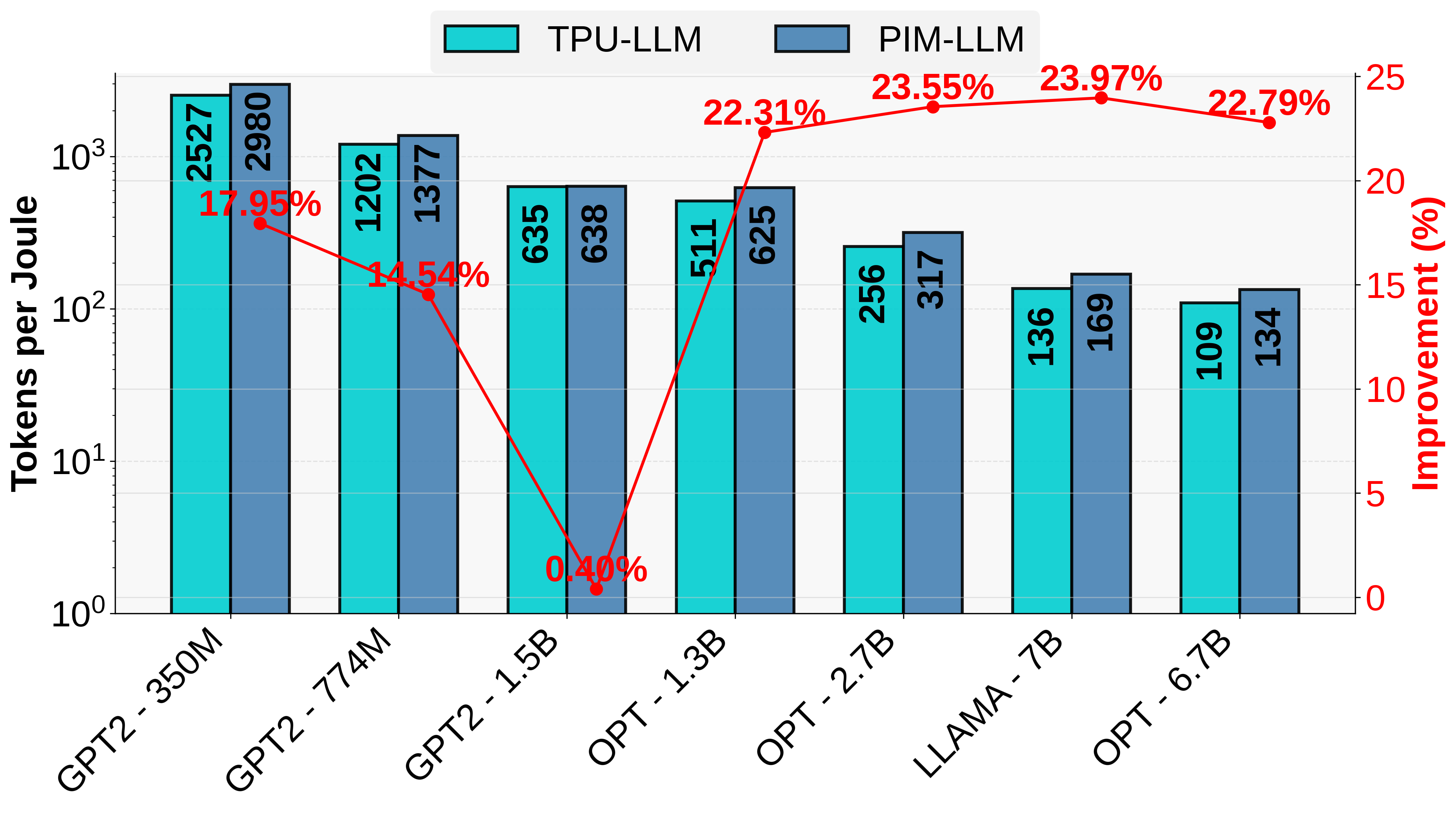}}
\hfill
\subfigure[$l=4096$]{\includegraphics[width=0.31\textwidth]{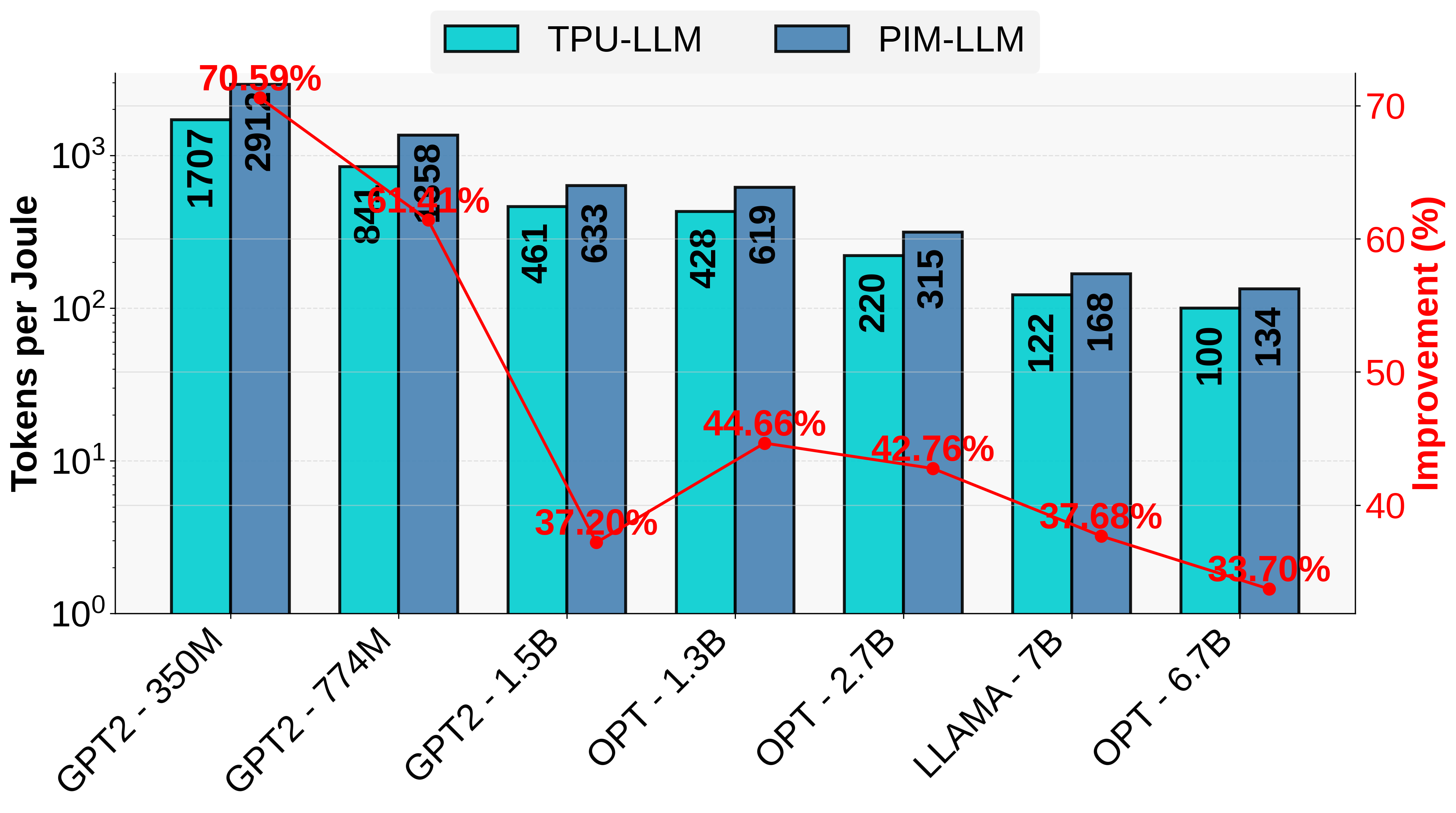}}
\hfill
\vspace{-3mm}
\caption{Tokens per Joule result for various LLMs with different context lengths ($l$).}
\label{fig:tokens_per_joule_improvement}
\end{figure*}

\subsection{Tokens per Second Analysis}
The proposed PIM-LLM Architecture architecture demonstrates significant performance gains compared to the TPU baseline. As shown in Fig. \ref{fig:tokens_sec_speedup}, the speedup varies across models and context lengths, with the most substantial gains observed for models with shorter context lengths. For instance, at a context length of 128, the GPT 350M model achieves an \textbf{11.6$\times$} speedup, while the OPT 6.7B model reaches a \textbf{79.2$\times$} speedup. This trend is consistent across different models, with larger models showing greater speedups, suggesting that the architecture will scale effectively for larger models.

However, as the context length increases, the speedup decreases. This is because, for larger context lengths, the MatMul operations in the attention heads become larger (refer to Table \ref{tab:dimension}), requiring more computation within the TPU component of the hybrid PIM-LLM architecture. It is worth noting that this issue is less significant for edge applications, as they generally operate with shorter context lengths \cite{chen2024octopus}. As illustrated in Fig. \ref{fig:tokens_sec_speedup}, for a context length of 4096, the GPT 350M model shows a \textbf{1.5$\times$} improvement, while the OPT 6.7B model achieves a still considerable \textbf{5.71$\times$} speedup. 



\subsection{Latency Percentage Breakdown}


Figure \ref{fig:latency} provides a detailed breakdown of latency contributors within the PIM-LLM architecture. The systolic arrays, which handle MatMul operations in the attention heads, represent the largest latency component. For the OPT 6.7B model, they account for \textbf{60\%} of total latency, increasing to \textbf{73.9\%} for the GPT-2 350M model at a context length of 128. This pattern holds across various model sizes. At a context length of 4096, the systolic arrays account for over \textbf{97\%} of latency in both the GPT-2 350M and OPT 6.7B models.

The latency contributed by the digital peripheral circuitry is negligible, accounting for less than \textbf{0.01\%}, and thus barely visible even in the zoomed sections of the graphs. As the context length increases, latency from the systolic arrays also rises, although it decreases as model size grows. The combined latency of RRAM crossbars (Xbar), DAC, and ADC remain below \textbf{1\%} highlighting the effectiveness of the PIM component to accelerate the projection layers in LLMs. Communication latency becomes more prominent in larger models with shorter context lengths, contributing \textbf{36.3\%} for the OPT 6.7B model and \textbf{10.7\%} for the GPT-2 350M model at a context length of 128. The buffer latency percentage is \textbf{14.7\%} and \textbf{3.5\%} for the GPT-2 350M and the OPT 6.7B model, respectively. For other context lengths, the buffer latency contribution ranges from \textbf{1\%} to \textbf{5\%}. The zoomed section highlights values below 1\%, typically reflecting PIM latency and, in some cases, buffer latency.

\begin{figure*}[t]
\centering
\subfigure[$l=128$]{\includegraphics[width=0.31\textwidth]{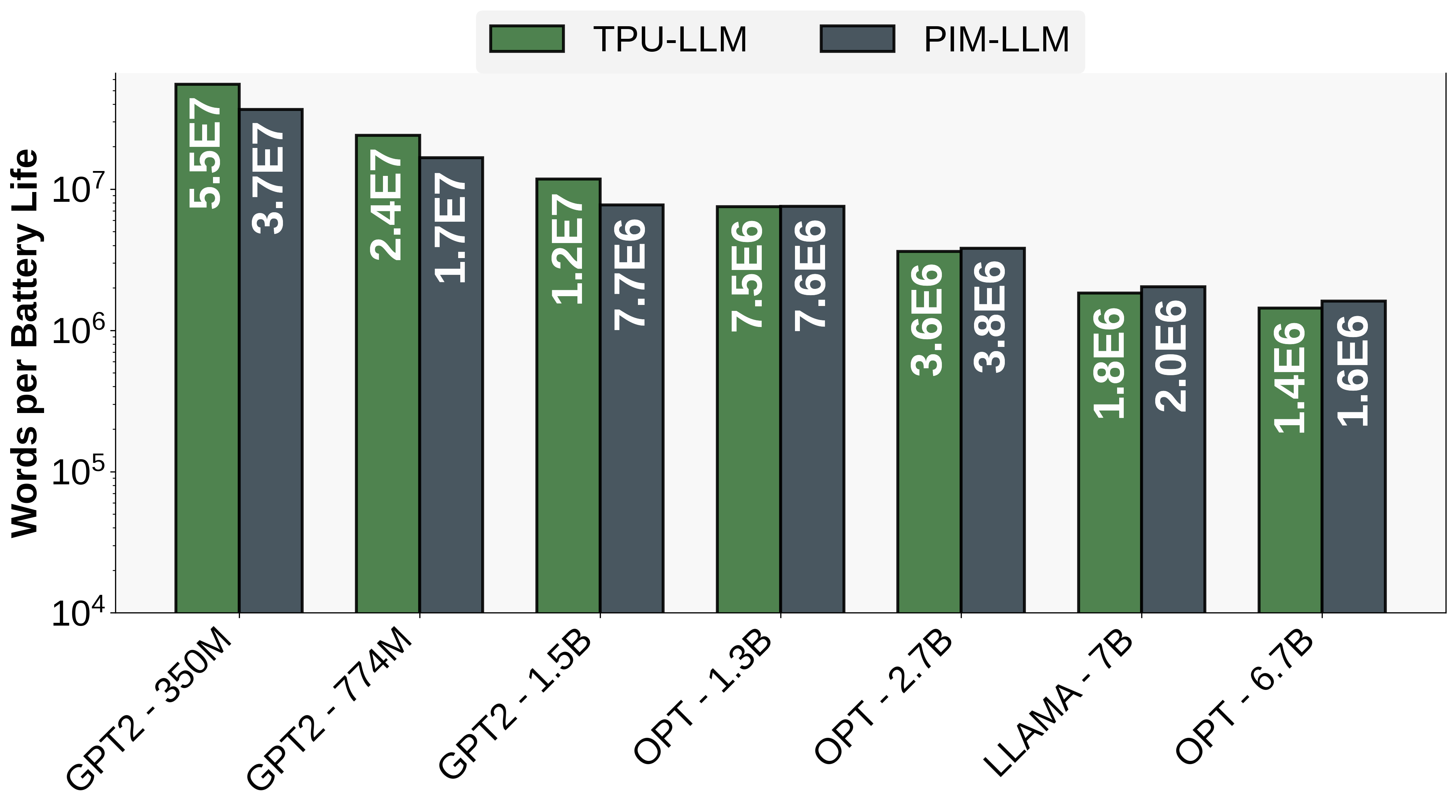}}
\hfill
\subfigure[$l=256$]{\includegraphics[width=0.31\textwidth]{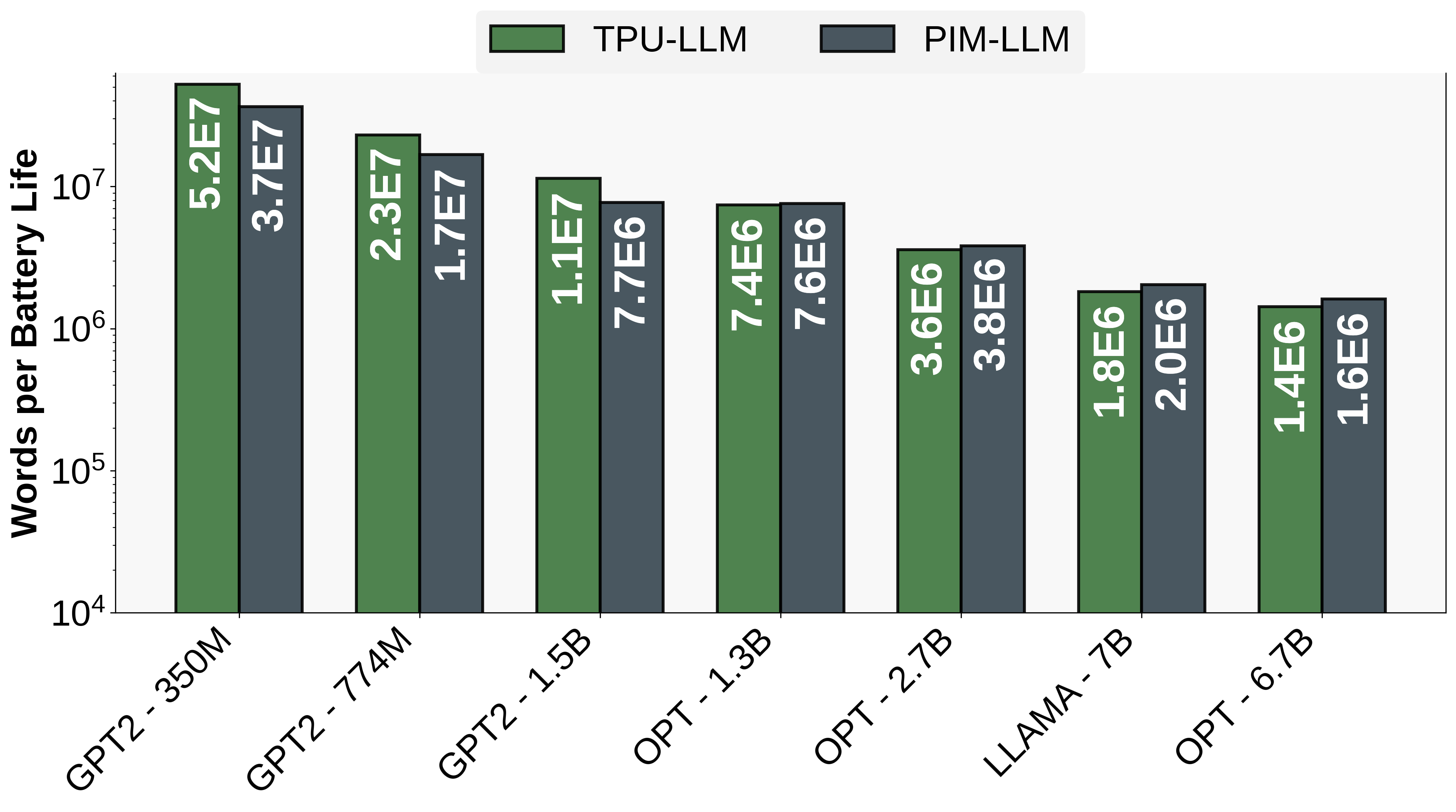}}
\hfill
\subfigure[$l=512$]{\includegraphics[width=0.31\textwidth]{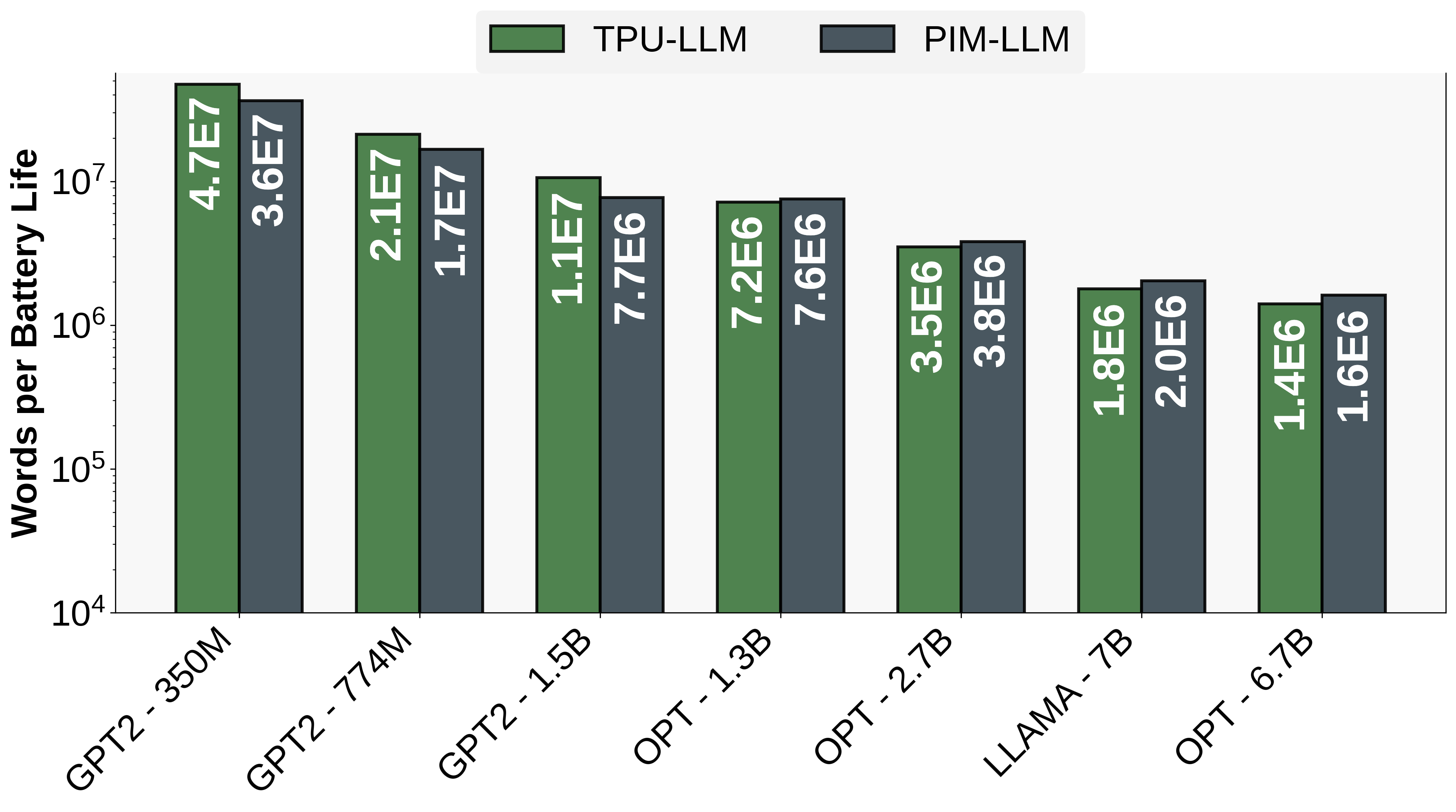}}
\hfill
\subfigure[$l=1024$]{\includegraphics[width=0.31\textwidth]{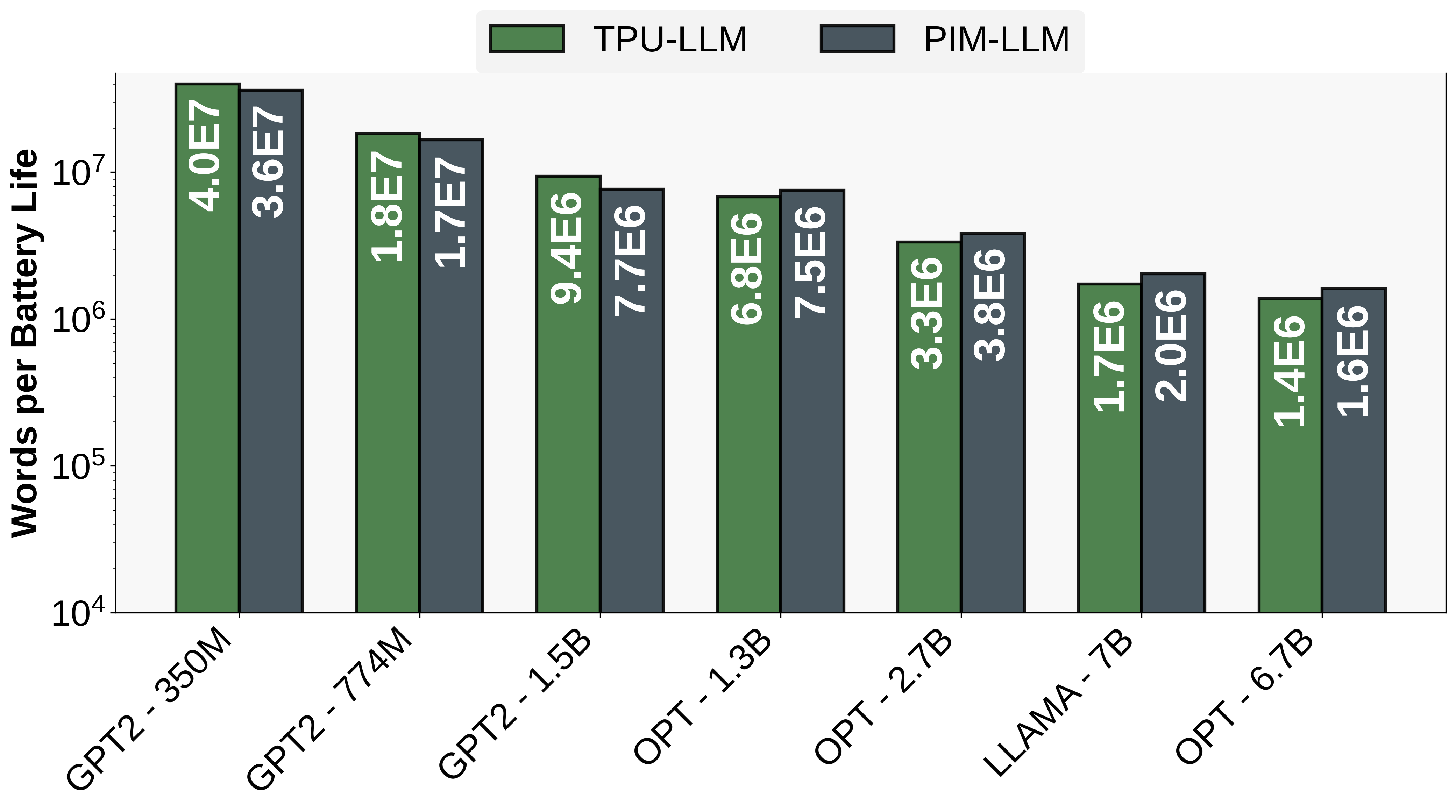}}
\hfill
\subfigure[$l=2048$]{\includegraphics[width=0.31\textwidth]{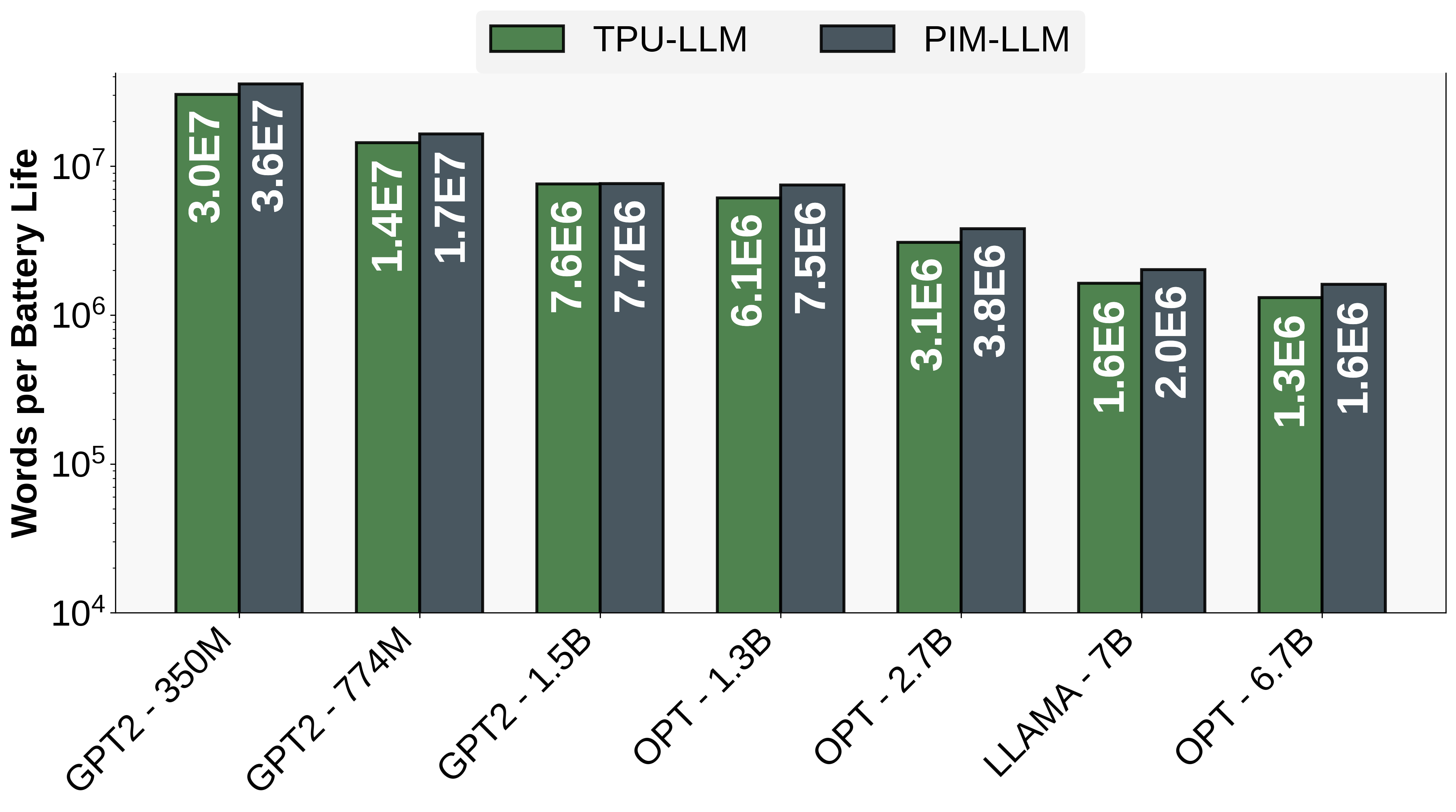}}
\hfill
\subfigure[$l=4096$]{\includegraphics[width=0.31\textwidth]{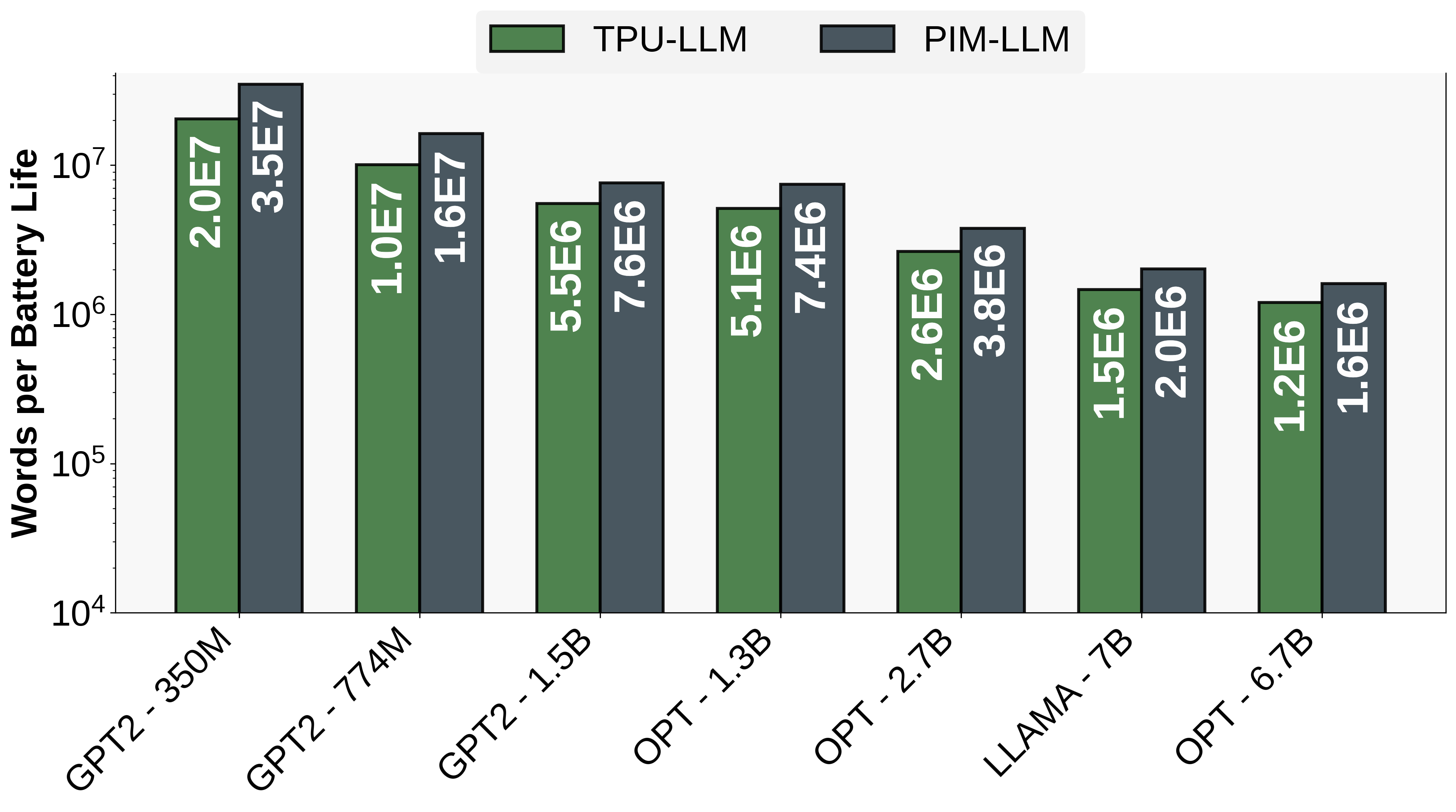}}
\hfill
\vspace{-2mm}
\caption{Words Per Battery Life result for various LLMs with different context lengths ($l$).}
\label{fig:words_per_battery_life}
\vspace{-2mm}
\end{figure*}

\subsection{Tokens per Joule Analysis}
We also analyze throughput per joule of energy consumed, as energy efficiency is a critical metric, particularly for energy-constrained edge devices. Figure \ref{fig:tokens_per_joule_improvement} highlights the improvement in tokens per joule across different context lengths. For smaller models and context lengths, the TPU outperforms the proposed PIM-LLM architecture, particularly with the GPT-2 350M model, where TPU delivers \textbf{33.7\%} lower energy consumption. This trend persists across context lengths of 256, 512, and 1024. This is due to the fact that the substantial speedup provided by the PIM architecture comes at the cost of high power dissipation.


However, once model sizes exceed the OPT 1.3B model, our PIM-LLM architecture begins to demonstrate gains. For instance, at a context length of 128 for the OPT 1.3B model, the proposed architecture is \textbf{0.96\%} more energy efficient, with improvements reaching \textbf{12.49\%} for the OPT 6.7B model. Notably, the results reveal a significant increase in efficiency as context length increases. At context lengths of 2048 and 4096, the proposed architecture consistently outperforms TPU-LLM across all model sizes, achieving gains of \textbf{17.95\%} for GPT-2 350M and \textbf{22.79\%} for OPT 6.7B at $l=2048$, and \textbf{70.58\%} for GPT-2 350M and \textbf{33.7\%} for OPT 6.7B at $l=4096$.


\subsection{Words per Battery Life}


Considering a standard edge battery with a capacity of 5 watt-hours, equivalent to 18,000 Joules, and assuming a conservative average of 1.5 tokens per word \cite{lambruschini2023reducing}, we introduce a new metric called \textit{Words/Battery Life}. The results of this metric are shown in Fig. \ref{fig:words_per_battery_life}. For OPT-6.7B with a context length of 128, PIM-LLM architecture can achieve \textbf{1.6M} words/battery life compared to \textbf{1.4M} words on TPU-LLM. As the context length increases to 1024, performance remains comparable across both TPU-LLM and PIM-LLM architectures. However, at context lengths of 2048 and 4096, PIM-LLM consistently outperforms TPU-LLM across all model sizes. For example, for $l=4096$, the GPT2-350M model reaches \textbf{35M} words, and OPT-6.7B achieves \textbf{1.6M} words on PIM-LLM, whereas TPU-LLM performance for these models is capped at \textbf{20M} and \textbf{1.2M} words, respectively.


\subsection{Related Work}
\begin{table}[]
\centering
\caption{Comparison with previous works.}
\begin{tabular}{clcc}
\hline
\multicolumn{1}{l}{}     & \multicolumn{1}{c}{Model} & GOPS & GOPS/W         \\ \hline
TransPIM \cite{TRANSPIM}                 & GPT2-Medium ($l=4096$)      & -    & \textless{}200 \\ 
HARDSEA \cite{Hardsea}                 & GPT2-Small ($l=1024$)       & 3.2  & -              \\ \hline
\multirow{2}{*}{PIM-LLM} & GPT2-Small ($l=1024$)       & \textbf{6.47} & 487.4          \\ 
                         & GPT2-Medium ($l=4096$)      & 3.7  & \textbf{1026}           \\ \hline
\end{tabular}
\label{tab:comparison}
\end{table}



Few prior studies have explicitly explored PIM-based architectures for decoder-only models, which dominate current generative language models. This focus is crucial, as works like HARDSEA \cite{Hardsea} and TransPIM \cite{TRANSPIM}—designed for both encoder-only and decoder-only models—report notable performance degradation when implementing decoder-only models. These challenges arise from unique characteristics of decoder-only models, such as their heavy reliance on matrix-vector multiplications to generate tokens sequentially, one at a time.

Moreover, the few studies that have focused on decoder-only models primarily examine smaller-scale models that are not typically classified as ``large'' language models. For instance, HARDSEA \cite{Hardsea} and TransPIM \cite{TRANSPIM} implement the GPT-2 Small and GPT-2 Medium models, respectively, which are smaller than the OPT and LLAMA models analyzed in this work. PIM-GPT \cite{PIM-GPT} is a recent work that examines large GPT models, however it only introduces light modifications to DRAM and still relies on digital computation with 16-bit floating-point number representation. Consequently, PIM-GPT is more comparable to large-scale digital cloud accelerators than to the proposed PIM-LLM architecture here, which is specifically tailored for edge applications.

In this work, we compare our approach with HARDSEA \cite{Hardsea} and TransPIM \cite{TRANSPIM}. Without access to the detailed implementations of these prior designs, we rely on the throughput values explicitly mentioned in their papers. As listed in Table \ref{tab:comparison}, TransPIM \cite{TRANSPIM} presents several variations of its architecture, all achieving a \textit{GOPS/W} below 200 on the GPT-2 Medium model with a 4096 context length. In contrast, HARDSEA reports a throughput of 3.2 GOPS when running GPT-2 Small with a 1024 context length. Using our proposed PIM-LLM architecture for the same workloads, we achieve more than a $5\times$ improvement in GOPS/W compared to TransPIM and a $2\times$ improvement in GOPS compared to HARDSEA. It is worth noting that PIM-LLM demonstrates even greater benefits with larger language models, so comparisons with prior work on small GPT models may not fully reflect its potential. For example, with the OPT 6.7B model and a context length of 1024, the GOPS and GOPS/W values increase to 58.5 and 1134.14, respectively. Similarly, for the OPT 6.7B model with a context length of 4096, the GOPS and GOPS/W values increase to 17.6 and 1262.72, respectively.

\renewcommand{\arraystretch}{1.1} 

\section{Conclusion}

The advent of 1-bit LLMs has created significant research opportunities to lower the computational demands of generative LLMs and enhance their throughput efficiency. Although some recent studies have investigated this area, there is still a lack of practical architectural implementations that capitalize on the unique properties of 1-bit LLMs. This paper introduced PIM-LLM, a hybrid analog-digital architecture that achieves considerable performance and throughput gains compared to conventional LLM accelerators and the limited prior efforts that have explicitly focused on smaller-scale decoder-only language models. Our work sets a new benchmark for accelerating decoder-only LLMs using PIM technology, achieving a scale and efficiency previously unattained.

\section{Acknowledgments}
This work is supported in part by the National Science Foundation (NSF) under grant numbers 2340249 and 2409697.

\balance
\bibliographystyle{IEEEtran}
\bibliography{Reference.bib}

\end{document}